\documentclass[fleqn,usenatbib]{mnras}
\usepackage{newtxtext,newtxmath}
\usepackage[T1]{fontenc}
\usepackage{graphicx}	%
\graphicspath{{./}{figures/}}
\usepackage{soul}
\usepackage{xcolor}
\usepackage[normalem]{ulem}
\usepackage{xspace}
\newcommand{\orcid}[1]{\href{https://orcid.org/#1}{\includegraphics[width=10pt]{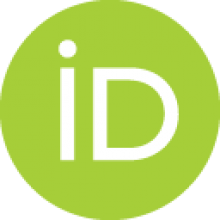}}}

\newcommand{\vvec}[1]{\mathbf{#1}}

\newcommand{\mpcph}{~h^{-1}{\rm Mpc}}

\newcommand{\weave}{\mbox{{\sc Weave}}\xspace}
\newcommand{\ns}{n_\mathrm{s}}

\newcommand{\Rs}{R_\mathrm{s}}
\newcommand{\Rex}{R_\mathrm{ex}}
\newcommand{\Om}{\Omega_{\rm m}}
\definecolor{bubbles}{rgb}{0.91, 1.0, 1.0}
\definecolor{aquamarine}{rgb}{0.5, 1.0, 0.83}
\definecolor{bubblegum}{rgb}{0.99, 0.76, 0.8}
\definecolor{bluebell}{rgb}{0.74, 0.74, 0.92}
\definecolor{purple}{rgb}{0.9,0.9,0.95}

\title[Probing cosmology via the clustering of critical points]{Probing cosmology via the clustering of critical points }
\author[J. Shim, C. Pichon, D. Pogosyan, S. Appleby, C. Cadiou, J. Kim \& K. Kraljic]{\parbox[t]{\textwidth}{
Junsup Shim$^{1}$\thanks{E-mail:jshim@asiaa.sinica.edu.tw}\orcid{0000-0001-7352-6175}, Christophe Pichon$^{2,3,4}$\orcid{0000-0003-0695-6735}, Dmitri Pogosyan$^5$\orcid{0000-0002-7998-6823},\\
Stephen Appleby$^{6,7}$\orcid{0000-0001-8227-9516}, Corentin Cadiou$^{8}$\orcid{0000-0003-2285-0332}, Juhan Kim$^{9}$\orcid{0000-0002-4391-2275}, Katarina Kraljic$^{10}$\orcid{0000-0001-6180-0245}, and Changbom Park$^{2}$\orcid{0000-0001-9521-6397} \\
}
\\
$^{1}$ Institute of Astronomy and Astrophysics, Academia Sinica, No.1, Sec. 4, Roosevelt Rd, Taipei 10617, Taiwan\\
$^{2}$ School of Physics, Korea Institute for Advanced Study, 85 Hoegi-ro, Dongdaemun-gu, Seoul, 02455, Republic of Korea\\
$^{3}$ CNRS and Sorbonne Universit\'e, UMR 7095, Institut d'Astrophysique de Paris, 98 bis Boulevard Arago, F-75014 Paris, France\\
$^{4}$ Université Paris-Saclay, CNRS, CEA, Institut de physique théorique, 91191, Gif-sur-Yvette, France.\\
$^{5}$ Department of Physics, University of Alberta, 11322-89 Avenue, Edmonton, Alberta, T6G 2G7, Canada.\\
$^{6}$ Asia Pacific Center for Theoretical Physics, Pohang, 37673, Republic of Korea\\
$^{7}$ Department of Physics, POSTECH, Pohang, 37673, Republic of Korea\\
$^{8}$ Lund Observatory, Division of Astrophysics, Department of Physics, Lund University, Box 43, SE-221 00 Lund, Sweden\\
$^{9}$ Center for Advanced Computation, Korea Institute for Advanced Study, 85 Hoegiro, Dongdaemun-gu, Seoul 02455, Republic of Korea\\
$^{10}$  Observatoire Astronomique de Strasbourg, Universit\'e de Strasbourg, CNRS, UMR 7550, F-67000 Strasbourg, France\\
}
\date{\today}

\pubyear{2023}

\begin{document}
\label{firstpage}
\pagerange{\pageref{firstpage}--\pageref{lastpage}}
\maketitle

\begin{abstract}
Exclusion zones in the cross-correlations between critical points (peak-void, peak-wall, filament-wall, filament-void) of the density field define quasi-standard rulers that can be used to constrain dark matter and dark energy cosmological parameters. 
The average size of the exclusion zone is found to scale linearly with the typical distance between extrema. The latter 
changes as a function of the matter content of the universe in a predictable manner, but its comoving size remains essentially constant in the linear regime of structure growth on large scales, unless the incorrect cosmology is assumed in the redshift-distance relation.
This can be used to constrain the dark energy parameters when considering a survey that scans a range of redshifts.
The precision of the parameter estimation is assessed using a set of cosmological simulations, and is found to be a 4$\sigma$ detection of a change in matter content of 5\%, or about 3.8$\sigma$ detection of 50\% shift in the dark energy parameter using a full sky survey up to redshift 0.5.
\end{abstract}



\section{Introduction}
The large-scale matter distribution is a valuable source of information because its clustering properties are sensitive to cosmology.
Indeed, stringent constraints have been set on cosmological parameters thanks to measurements of the baryonic acoustic oscillation \citep[e.g.][]{eisenstein_DetectionBaryonAcoustic_2005a, 2007MNRAS.381.1053P, okumura+08, 2011MNRAS.416.3017B, dawson+13, 2021PhRvD.103h3533A, xu+23}, redshift-space distortions \cite[e.g.][]{2005MNRAS.361..879D,2010Natur.468..539M, okumra+16, neveux+20}, and Alcock-Paczynski effect \cite[e.g.][]{blake+11,li+16, beutler+17, li+18, zhang+19, dong+23} in the two-point correlation functions of galaxies. However, as the evolution of the density field becomes increasingly non-linear, it departs from its Gaussian initial state. This causes an increasing amount of information to be contained in statistics beyond the two-point functions, which can be captured by measuring three-point correlation functions \citep[e.g.][]{Peebles+1975,Hinshaw+1995, Nichol+2006,marin+13,slepian+17,sugiyama+23} or higher-order moments
\citep[e.g.][]{Fry1985,Bouchet+1993,Bernardeau1994,Croton+2004,Cappi+2015,sabiu+19,philcox22}.
While the hierarchy typically converges, each extra order becomes increasingly more difficult to measure robustly.
This has fostered the development of alternative probes to obtain information beyond the simple two-point functions.

One avenue is to weigh the tracers used to compute the correlation functions according to their properties -- such as the local density or galaxy properties -- to obtain so-called marked statistics \citep[][]{2009MNRAS.395.2381W,Uhlemann2017,armijo+18,satpathy+19,massara+21, massara+23} or revert back to using one-point statistics \citep[e.g.][]{2000A&A...364....1B,uhlemann+16,2020MNRAS.495.4006U,2021MNRAS.503.5204B,2023OJAp....6E..22B,marques+23}.

On large scales, one of the most striking features of the matter distribution is the presence of the cosmic web, composed of voids, walls in between them, separated by filaments which finally intersect at cosmic nodes \citep{bond+96}.
This has sparked interest in building alternative probes informed by the topology of the cosmic web to measure cosmological parameters, such as
the genus curve and Euler-Poincar\'e characteristic \citep[e.g.][]{GottMellotDickinson1986,Melott+1989,ParkGott1991,MeckeWagner1991,Park+2001,BerianJames+2009,ParkKim2010,appleby+18,appleby+21}, Minkowski functionals \citep[e.g.][]{MeckeBuchertWagner1994,SchmalzingBuchert1997,hikage+03,natoli+10,junaid+15,appleby+18minkowski_1,goyal+20minkowski,appleby+22}, percolation  \citep[e.g.][]{Shandarin1983,Yess+1997,ShandarinYess1998,Colombi2000,zhang+18} and skeleton \citep[e.g.][]{sousbie+08,sousbie+09,Sousbie2011-2}, or persistent homology analysis \citep[e.g.][]{,Sousbie2011-2,Pranav+2017}, alpha-shapes and Betti numbers \citep[e.g.][]{vandeWeygaert+2011,chingangbam+12,Park+2013,Pranav+2019,Feldbrugge+2019}.

Recently, it has been shown that the relative clustering of critical points of a density field is maintained nearly-constant throughout the gravitational evolution \citep[][Appendix B]{shim+21}, and can be measured accurately.
Critical points are topological elements of a given field, and their attributes including position, height, curvature, and relative orientation encode the topological characteristics of the underlying field \citep{bond+96,pogosyan+09,sousbie+09,Gay2012} and its evolution \citep{cadiou2020}. One of their characteristic clustering features is that a pair of critical points with different curvatures and heights cannot be arbitrarily close. This exclusion zone, or the strong anti-clustering region \citep{lumsden+89,mo&white96,sheth&lemson99,baldauf+16,shim+21}, appears more evidently in the cross-correlations between critical points with the opposite-sign biases \citep{shim+21}. Interestingly, \cite{shim+21} showed that the sizes of the exclusion zones in the initial Gaussian field are fairly consistent with those measured at late time, suggesting that we have a theoretical handle on their cosmology dependence since the Gaussian expectation value can be derived from first principles. On the other hand, the amplitude of this exclusion zone is shown to depend on how smooth the underlying field is \citep{baldauf+16}, indicating a cosmological dependence which we set forth to establish in this paper.

In expanding such exploration into cosmological tests, the redshift invariance of topology statistics emerges as a pivotal metric, as first introduced in \citet{ParkKim2010}. Leveraging the conserved nature of the genus amplitude in density fields smoothed on large scales, these authors present a method to constrain the cosmological model. This involves identifying the correct expansion history of the Universe that minimizes the evolution of the genus amplitude with redshift. Aligning with this strategy, we introduce an approach utilizing the exclusion radius as a standard ruler to probe cosmology without delving into specific practicalities. We measure the size of the exclusion zone in critical point correlation functions and show how it can perform as a cosmological probe. Specifically, we focus on cross-correlations involving peak-wall, peak-void, filament-wall, and filament-void pairs -- tracers that are oppositely biased to matter density fields. We rely on the suite of cosmological $N$-body simulations to measure the critical point statistics.

The outline of the paper is as follows: 
  \S\ref{sec:simu}  describes briefly   the multiverse simulation;
  \S\ref{sec:estimator} introduces estimators for the exclusion zone;
  \S\ref{sec:results} presents our results on the $\Omega_\mathrm{m}$-dependent variations of the exclusion zone  radius,
  and on estimating  $w_{\rm de}$  when different redshifts fields are considered,
  while
  \S\ref{sec:conclusions} concludes.
  Appendix~\S\ref{sec:appendix-16} shows the apparent evolution of the exclusion radius for a larger smoothing scale, while \S\ref{sec:appendix-scaling} discusses the theoretical expectation of the exclusion radius.

\section{Simulation set}\label{sec:simu}

In this paper, we rely on the multiverse simulations introduced in \citet{park+19, tonegawa+20} which are a set of cosmological $N-$body simulations designed to test the effects of cosmological parameters on the clustering  of cosmic
structures. 
This set of five simulations varies the cosmological parameters centered on a fiducial $\Lambda$CDM cosmology with $\Om=0.26$, $\Omega_\mathrm{de}=0.74$, $H_{0}=72$km/s/Mpc, and $w_{\rm de}\equiv p_{\rm de}/\rho_{\rm de}=-1$, \citep[in alignment with WMAP5 constraints;][]{dunkley+09}. The initial displacement and velocity of the simulated dark matter particles are obtained by applying the second-order Lagrangian perturbation theory \citep{jenkins+10}. The same set of random numbers has been employed to produce the initial density perturbations across all simulations, enabling a more straightforward comparison between them without the confounding effects of cosmic variance. Specifically, two of these simulations involve an alteration in the matter density parameter by $\pm 0.05$ relative to the fiducial model, maintaining the dark energy equation of state (EOS) at $w_{\rm de}=-1$. The remaining two simulations, based on quintessence models \citep{sefusatti&vernizzi11}, introduce a deviation in $w_{\rm de}$ by $\pm 0.5$ from the fiducial dark energy EOS, with $\Om$ consistently set to $0.26$. Cosmological parameters for the five cosmologies including the fiducial one are summarized in Table~\ref{tab:cospar}.

\begin{table}
\centering
\caption{Summary of cosmological parameter values of five cosmologies considered. The second-row model is the fiducial cosmology.}
\label{tab:cospar}	
    \begin{tabular}{cccc} \hline\hline
    & $\Om$ & $w_{\rm de}$ & $\Omega_{\rm de}$\\
    \hline
    ($\Om^{-}, w_{\rm de}^{0}$) & 0.21 &  -1.0 & 0.79\\[3pt]
    ($\Om^{0}, w_{\rm de}^{0}$) & 0.26 &  -1.0 & 0.74\\[3pt]
    ($\Om^{+}, w_{\rm de}^{0}$) & 0.31 &  -1.0 & 0.69\\[3pt]
    ($\Om^{0}, w_{\rm de}^{-}$) & 0.26 &  -1.5 & 0.74\\[3pt]
    ($\Om^{0}, w_{\rm de}^{+}$) & 0.26 &  -0.5 & 0.74\\
    \hline
    \end{tabular}
\end{table}

The power spectra are normalized such that the root mean square of the linearly evolved matter fluctuation at $z=0$ yields a value of $\sigma_{8}=0.794$ when smoothed with a spherical top hat
with $R_{\rm s}=8\,\mpcph$.
The number of particles in each simulation is $2048^3$ and the
comoving size of the simulation box is $1024\,\mpcph$. The initial power spectrum at a redshift $z_{\rm init}=99$ was computed with the CAMB package. The  N-body integrator is  an extension of the original GOTPM code
\citep{dubinski+04}
which evolve particles according to the modified Poisson equation 
\begin{equation}
    \nabla^2\phi = 4\pi Ga^2 \bar{\rho}_\mathrm{m} \delta_\mathrm{m}\left(1 +\frac{D_\mathrm{de}}{D_\mathrm{m}}\frac{\Omega_\mathrm{de}(a)}{\Omega_\mathrm{m}(a)}\right),
\end{equation}
where $D_\mathrm{de}$ and $D_\mathrm{m}$ are the linear growth factors of the dark energy and matter, respectively (see \citealt{sefusatti&vernizzi11} for details).

\begin{figure*}
\includegraphics[width=2\columnwidth]{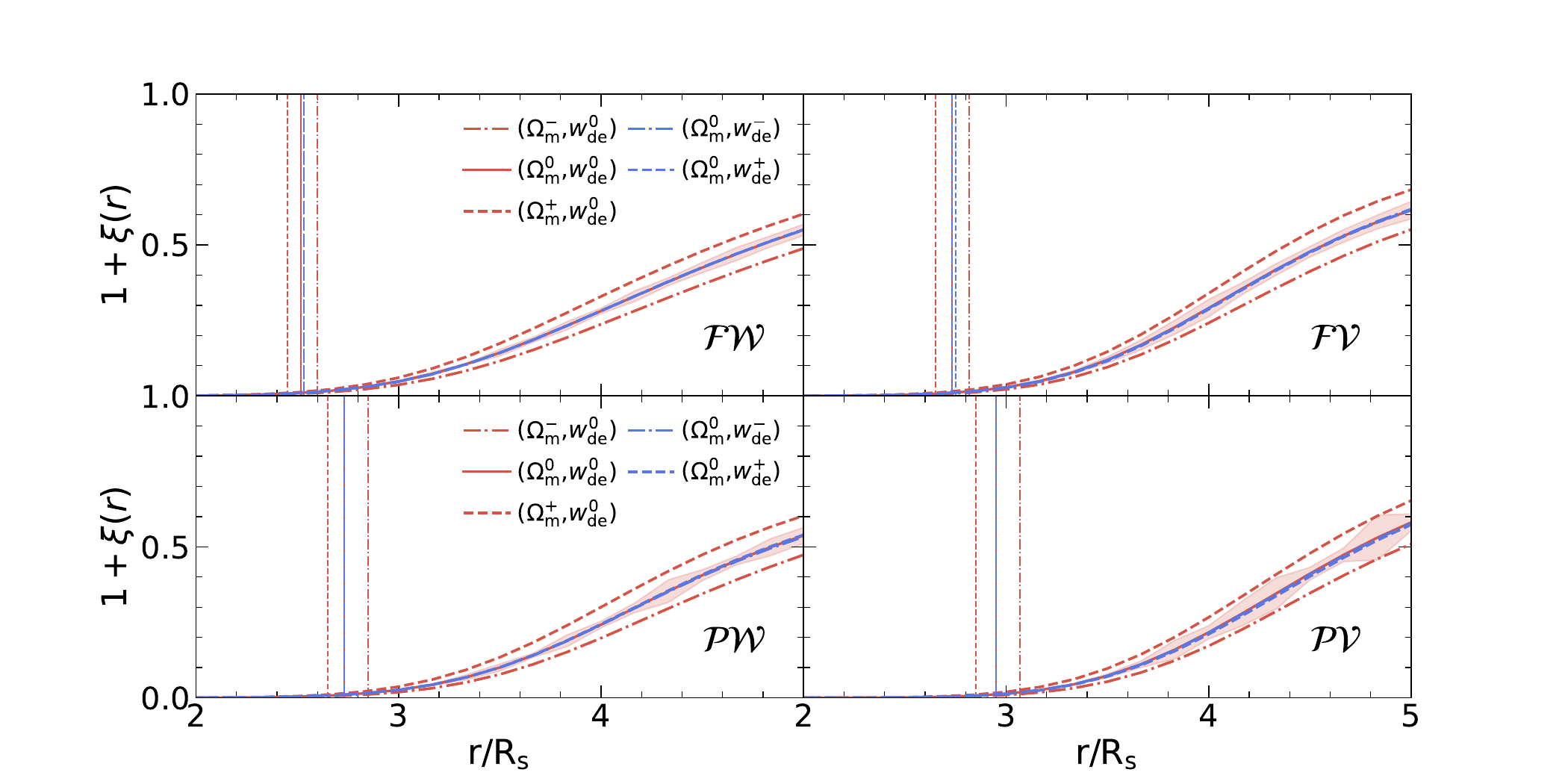}
    \caption{Two-point cross-correlation functions for pairs of positively and negatively biased critical points for the five different cosmologies as labeled. Correlation functions for the ${\cal FW}$, ${\cal FV}$, ${\cal PV}$, and ${\cal PW}$ are shown  clockwise  from the upper-left panel. The adopted Gaussian smoothing scale $R_{\rm s}$ is $6\mpcph$. 
    Vertical lines represent the exclusion zone radii and shaded regions show the standard errors around the fiducial cosmology. Note that the correlation function for the fiducial cosmology (red-solid) is nearly identical to those with the different equations of state dark energy models (blue). 
    However, the two-point correlation function depends on $\Omega_\mathrm{m}$, see Fig.~\ref{fig:Rex_cosmo}. 
    }
    \label{fig:Cross2pt}
\end{figure*}

\begin{figure*}
\includegraphics[clip,width=1.6\columnwidth]{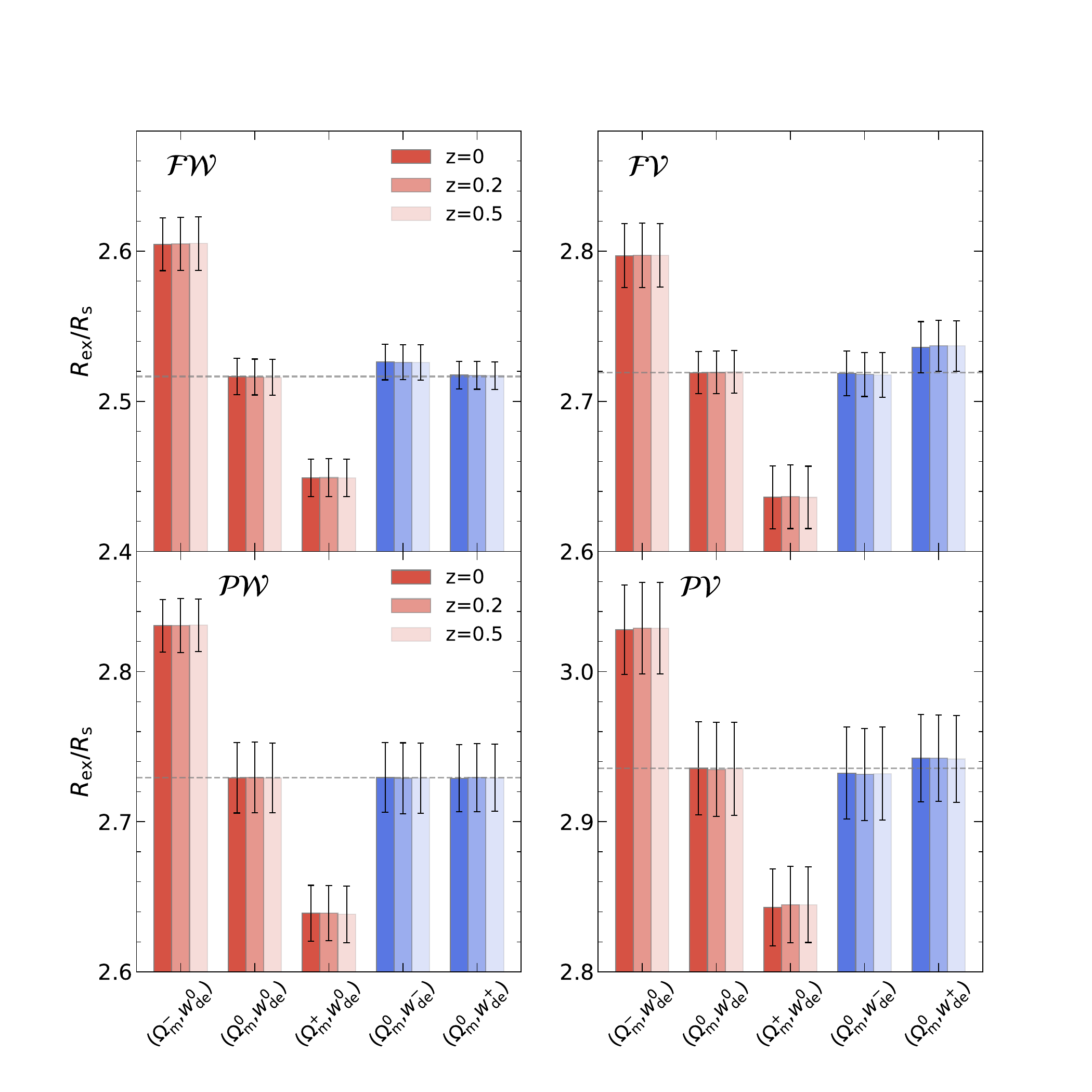}
\caption{
Exclusion zone radius averaged over 8 measurements from the simulations for the five different cosmologies. Different $\Om$-- and $w_{\rm de}$-- cosmologies are shown in red and blue, respectively. We normalized the exclusion zone radius by the smoothing scale $R_{\rm s}$ adopted since $\Rex$ linearly depends on the smoothing scale. Error bars represent the standard errors of the mean. The grey horizontal line marks the exclusion zone radius for the fiducial cosmology with $\Om=0.26$ and $w_{\rm de}=-1$. For this plot, we adopt the Gaussian smoothing scale $R_{\rm s}=6\mpcph$. 
The trend of the exclusion radius with $\Om$ is consistent with the scaling involving $n_s$ in equation~\eqref{eq:defR*PL} if one associates $\Rex$ with $R_\ast$, see the text for details.
}
\label{fig:Rex_cosmo}
\end{figure*}

\begin{figure*}
\includegraphics[clip,width=1.8\columnwidth]{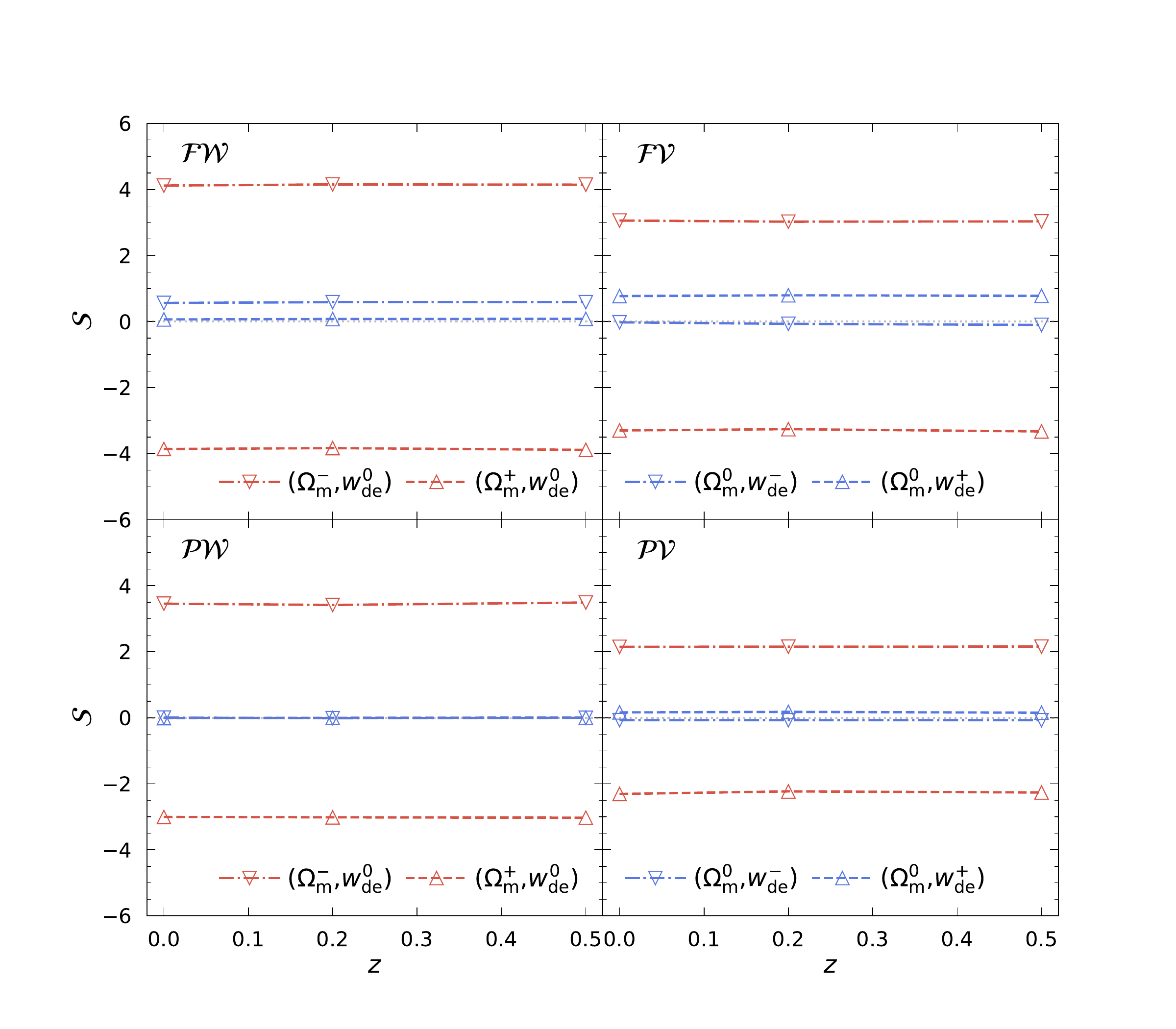}
\caption{Significance, $\cal S$, of the difference in the exclusion radius from the fiducial cosmology as a function of redshift. Non-fiducial $\Om$ and $w_{\rm de}$ models are shown in red and blue, respectively. For this plot, we adopt the Gaussian smoothing scale $\Rs=6\mpcph$. Note that we allow the significance to have both signs as they can indicate the direction of the difference from the fiducial case.}
\label{fig:rex_signi}
\end{figure*}

\begin{figure*}
\includegraphics[clip,width=1.6\columnwidth]{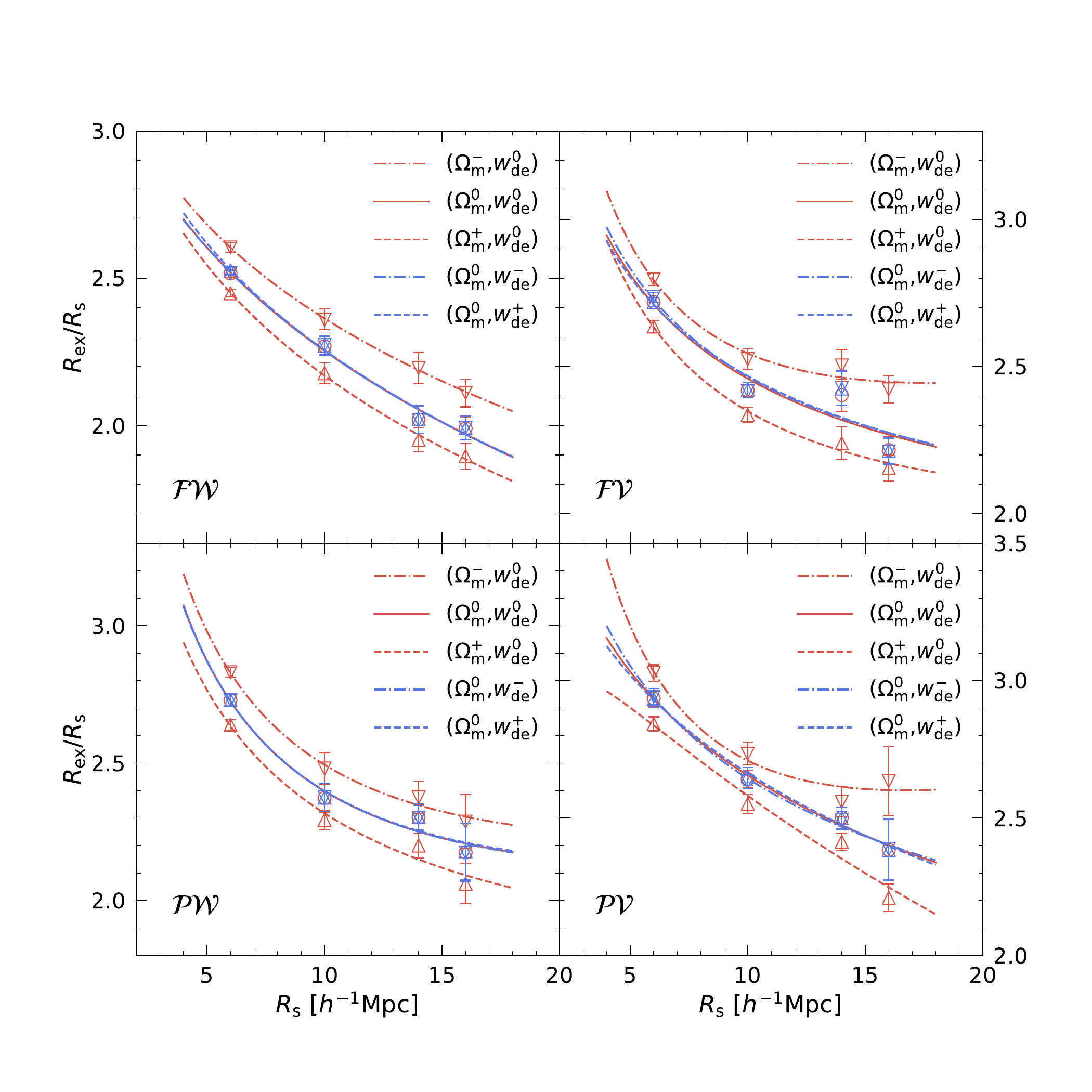}\caption{Measured exclusion zone radius as a function of smoothing scale for four different types of cross-correlations. Correlation functions for ${\cal FW}$, ${\cal FV}$, ${\cal PV}$, and ${\cal PW}$ are shown clockwise from the upper-left panel. The symbols and lines represent the measured exclusion radius and best fit to these measurements, respectively. Error bars are measured as the standard error of the mean.
Again the trend is consistent with theoretical expectation if one associates $\Rex$ to $R_\ast$, as confirmed to the first order in Fig.~\ref{fig:Rstar_Rex}.
}
\label{fig:Rex_Rs_fit}
\end{figure*}

\section{Exclusion zone estimators} \label{sec:estimator}

Let us first briefly describe how we define and measure the size of the exclusion zones. 
We refer to \cite{shim+21} for more details.
The first step is to compute a smooth density field in real space at the relevant redshifts. Density fields are calculated on $512^3$ grids by applying the Cloud-In-Cell method to the dark matter particle distribution. We then smooth these density fields with Gaussian kernel over 4 different smoothing scales $R_{\rm s}=16, 14, 10$ and $6\,\mpcph$. Note that the smallest smoothing scale corresponds to a typical size of an average density region of mass-scale around $10^{15} M_\odot$.

\subsection{Critical points}
The critical points are defined as positions where the  gradient of the dark matter density field vanishes \citep{milnor1963morse,BBKS,shim+21}. 
Based on the typical shape of isosurfaces in that neighborhood,  the four types are labeled 
according to the sign of the Hessian's eigenvalues: peaks (${\cal P}$) with signature
{\tiny ${-\!-\!-}$}, voids (${\cal V}$) with signature {\tiny ${+\!+\!+}$}, filament-type saddles (${\cal F}$) with signature {\tiny ${-\!-\!+}$}, and wall-type saddles 
(${\cal W }$) with signature {\tiny ${-\!+\!+}$}. They are proxies for the geometry of the cosmic web \citep{bond+96} of the underlying density field, tracing respectively 
clusters, voids, filaments, and walls \citep{pogosyan+09, sousbie+09} on a mass scale relevant to the adopted smoothing scale.

\subsection{Finding critical points}
\label{sec:algo}
The detection of  critical points in a smoothed dark matter density field relies on a second-order Taylor-expansion of the density field near  a critical point, $\vvec{x_c} $:
\begin{equation}
    \vvec{ x}-\vvec{x_c} \approx (\nabla\nabla \rho)^{-1} \nabla \rho ,
    \label{eq:finding_crit_point}
\end{equation}
where $\rho$ is the density field, $\nabla \rho$ the local gradient and $\nabla\nabla \rho$ the local hessian matrix.
The detection algorithm 
proceeds  as follows:
\emph{a)} For each cell in the grid, compute $\nabla \rho$ and $\nabla\nabla \rho$,
\emph{b)} solve equation~\eqref{eq:finding_crit_point}  discarding  solutions beyond a distance larger than one pixel and
\emph{c)} loop over cells that contain multiple critical points of the same kind, retaining for each only the critical point closest to the center of the cell \citep[the technique was originally introduced by one of the authors for][]{Colombi2000}.

\subsection{Computing  clustering correlation functions}

We count the pairs of critical points with rarity above or below a certain threshold to quantify their clustering characteristics. We define the rarity of critical points as 
\begin{equation}
    \nu\equiv{\delta}/{\sigma}, \quad \mathrm{with}\quad \delta\equiv{\rho}/{\bar{\rho}}-1, \quad \mathrm{and} \quad   \sigma^{2}\equiv\big\langle \delta^2\big\rangle \,,
\end{equation}
where $\delta$ is the over-density contrast of the smoothed density field, $\bar\rho$ the average density 
and $\sigma$ is the root mean square fluctuation of the field.

For peak and filament (respectively void and wall) critical points, we identify points with rarity higher (respectively lower) than a given threshold. In this analysis, the rarities $\nu^+_{\rm type}$ and $\nu^-_{\rm type}$ are chosen to trace the highest and lowest 20\%-rarity critical points, so that 
\begin{equation}
    \label{eq:rareness}
    N_{\rm type}(\nu \ge \nu^+_{\rm type})=0.2{N_{\rm type}},%
\quad
  N_{\rm type}(\nu \le \nu^-_{\rm type})=0.2{N_{\rm type}},%
\end{equation}
respectively for peaks and filaments, and  
for voids and walls. Here $N_{\rm type}(\nu \ge \nu_\mathrm{thresh} )$ represents the number of critical points of a given type
above the threshold $\nu_\mathrm{thresh}$ (here 20\% rarest) while $N_{\rm type}$ is the total number of critical points of this type. 
As discussed in \cite{shim+21}, this choice is driven by our requirement to sample populations that represent the same abundance for a given type of critical point. 
We  then measure cross-correlation functions with the estimator  given by \citet{1983ApJ...267..465D}
\begin{equation}
1+\xi_{ij}(r) = \frac{\langle C_{i}C_{j}\rangle}{\sqrt{\langle C_{i}R_{j}\rangle \langle C_{j}R_{i}\rangle }}
\sqrt{\frac{N_{R_i} N_{R_j}}{N_{C_i} N_{C_j}}},
\label{eq:xi}    
\end{equation}
where $C_i$ refers to a particular catalog $i\in\{{\cal P},{\cal F},{\cal W },{\cal V}\}$ and $R_i$ is a corresponding catalog with randomly uniformly distributed points in the same volume.  Here $\langle XY \rangle$ represents the number counts of the pairs between $X$ and $Y$  separated by $r$. The size of the sample, $N_{R_i}$, of the random catalog is a factor of $100$ or larger than the size of our simulated datasets, $N_{C_i}$.

\subsection{Defining exclusion radii}
The exclusion zone size, denoted as $\Rex$, is measured from the cross-correlation functions. Specifically, this is achieved by identifying the minimum distance at which the cross-correlation deviates from $\xi=-1$. In practice, we identify the radius at which
\begin{equation}
    1+\xi(\Rex)=\epsilon,
\label{eq:defRex}    
\end{equation}
with the deviation from perfect anti-correlation set at $\epsilon=0.01$. We note that the exclusion radius for a particular density field depends on the choice of rarity levels of critical points and on the choice of 
$\epsilon$ that defines departure from perfect anti-correlation. Decreasing rarity levels tends to yield smaller exclusion radii, as the difference in height between critical points will decrease. Conversely, it should increase the number of critical point pairs, thereby providing more reliable estimates for the exclusion radius.  

We estimate measurement uncertainties to evaluate the statistical significance of the impact of cosmological parameters on the exclusion radius. We divide the simulation box into eight separate, non-intersecting regions. For each sub-volume, we compute the cross-correlation function using critical points within that sub-volume. While doing this, we accounted for edge effects by creating a random sample within the same sub-volume. We then measure the exclusion radii from those cross-correlation functions. The measurement uncertainties are represented by the standard error of the mean, which is the standard deviation divided by the square root of the number of sub-volumes.

\section{Results}\label{sec:results}

Let us now turn to our main results on matter density and 
dark energy equation of state estimation using exclusion zone measurements.

\subsection{Exclusion radius and matter density parameter}\label{sec:results-omega}
We first explore how the exclusion radius varies with the matter density parameter. As illustrated in Fig.~\ref{fig:Cross2pt}, we measure the cross-correlation functions between peak/filament and void/wall critical points across five different cosmologies. The cross-correlations reveal varying exclusion zones based on the combination of critical points.
Notably, the filament-wall combination presents the smallest exclusion zone, while the peak-void combination showcases the largest. This observation aligns with the understanding that the exclusion zone expands with increasing differences in height and curvature, as shown in \citep{baldauf+16, shim+21}. After the emergence of the exclusion zone, cross-correlations start to deviate from $\xi=-1$ and become less anti-clustered with increasing separation.
Because peak/filament and void/wall points are oppositely biased tracers of the underlying matter density field, their cross-correlations are always negative, eventually approaching $\xi=0$ as expected \citep{Kaiser1984} at separations larger than $r\approx 10R_{\rm s}$.

Comparing cases for different matter density parameters reveals that the cross-correlations show a smaller exclusion radius for a larger matter density parameter. This implies that the exclusion radius is impacted by the matter density. In contrast, when focusing on dark energy models with non-standard dark energy EOS parameters $w_{\rm de}$, their exclusion radii are very similar to the fiducial model. This suggests that changes to the dark energy parameter do not significantly affect the exclusion radius. Thus, the size of the exclusion zone is mainly dictated by the amount of matter and remains largely unaffected by variations in the dark energy EOS.

We now quantitatively compare the mean exclusion radii across five different cosmologies at lower redshifts, as illustrated in Fig.~\ref{fig:Rex_cosmo}. The cross-correlations involving filaments and peaks are presented in the top and lower panels, respectively. Different shaded bars represent different redshift snapshot boxes for each cosmological parameter set. We observe a clear $\Om$-dependence of the exclusion radius. For instance, the exclusion zone shrinks as the matter density increases. However, adjusting the dark energy parameter shows a minimal impact on the exclusion radius. 
This is a consequence of the power spectrum slope at the scales of interest becoming shallower as $\Om$ increases but remaining nearly invariant with $w_{\rm de}$. We discuss the cosmological parameter dependence of the exclusion zone in detail below while relating it to the theoretical expectation for the distance between extrema.
Finally, when focusing on the time evolution, the exclusion zone remains remarkably stable in the redshift ranges investigated.

In Fig.~\ref{fig:rex_signi}, we quantify how significant the difference in the exclusion radius of each cosmology is from the fiducial case. The significance $\mathcal{S}$ is calculated as
\begin{equation}
\mathcal{S}\equiv\frac{R_{\rm ex}^{\rm x}-R_{\rm ex}^{\rm fid}}{\sqrt{\Sigma^{2}_{\rm x}+\Sigma^{2}_{\rm fid}}},
\end{equation}
where $R_{\rm ex}^{\rm x}$ and $\Sigma_{\rm x}$ represent the exclusion radius and standard error for a particular cosmology, while $R_{\rm ex}^{\rm fid}$ and $\Sigma_{\rm fid}$ are for the fiducial model. We find that the significance values tend to be larger for the filament-wall correlation compared to other critical point combinations. Conversely, the peak-void correlation shows the smallest significance.
This trend of larger (smaller) significance values when involving only saddles (extrema) arises from the differing numbers of critical points of each type. In a Gaussian random field, the ratio of saddle-to-extrema is approximately 3 \citep[see e.g.][]{Gay2012,shim+21}.
Consequently, the number of pairs for the filament-wall correlation is roughly an order of magnitude larger than that for the peak-void case. Therefore, the standard error is typically smaller for the case only involving saddles than extrema leading to a larger significance value for cross-correlations involving saddles. While the significance for one type of cross-correlation might not be sufficient to distinguish between different cosmologies, combining results from all four distinct measurements will enhance the overall constraining power.

We identify critical points on a particular scale determined by the smoothing length adopted, so next we examine how the exclusion radius changes with the smoothing scale, as this allows us to probe the clustering of critical points corresponding to a different mass scale. Fig. \ref{fig:Rex_Rs_fit} illustrates the variation of the exclusion radius measured from the matter density field smoothed on different smoothing scales. We observe a common trend that the rescaled exclusion radius decreases with the smoothing scale. This is evident across different cosmologies and types of cross-correlations considered. We consistently confirm that the exclusion radius shows a distinct dependence on the matter density on different smoothing scales. For instance, the rescaled $\Rex$ is always larger at all smoothing scales investigated for a cosmology with a smaller matter density. On the other hand, when varying the dark energy EOS parameters, the exclusion radii are remarkably consistent with the fiducial case.

Interestingly, we find that the behavior of $\Rex/\Rs$ seen in Fig.~\ref{fig:Rex_Rs_fit} is similar to the prediction for $R_{\ast}/R_{\rm s}$ as depicted in Fig.~\ref{fig:Rstar_Rs_theory}, which leads us to examine the relation between them. Here, $R_{\ast}$ represents the typical separation between extrema points and is defined by
\begin{equation}
R_\ast=\frac{\sigma_1}{\sigma_2}.
\label{eq:defR*}
\end{equation}
The moments of the smoothed power spectrum are calculated as
\begin{equation}
\sigma_{i}^{2}( R_{\rm s})\equiv\frac{1}{2\pi^2}\int^{\infty}_{0}\mathrm{d}k k^{2}P(k)k^{2i}W^{2}(k R_{\rm s})\,, \label{eq:defsig}
\end{equation}
where we adopt the Gaussian smoothing kernel,
\begin{equation}
    W(k R_{\rm s})=\exp\Big(-\frac{1}{2} k^2 R_{\rm s}^2\Big)\,,
\end{equation}
which serves as a low-pass filter suppressing power on scales below $1/\Rs$ for the power spectrum $P(k)$ weighted by varying powers of $k^{2i}$.
As depicted in Fig.~\ref{fig:Rstar_Rex}, we find a linear scaling relation between $\Rex$ and $R_{\ast}$. Associating these two provides a way to connect the measurements of exclusion radius to theoretical predictions based on Gaussian random fields. Indeed, for a Gaussian random field with (locally) scale-invariant power-spectra,  $P(k)\propto k^{\ns}$, the rescaled typical distance between extrema,
\begin{equation}
    \frac{R_\ast}{\Rs}=\sqrt{\frac{2}{\ns+5}} \label{eq:defR*PL}
\end{equation} 
is solely determined by the effective power-law index of the power spectrum, ${\ns}[k=1/\Rs]$. The effective slope is itself sensitive to the chosen smoothing scale, $\Rs$, and to the matter density, $\Om$, as displayed in Fig.~\ref{fig:5cosmoPS}, bottom panel.
Now, we can understand how the rescaled $R_{\ast}$ varies with respective matter density, dark energy EOS, and smoothing scale. Since the power spectrum slope either increases when the smoothing scale or matter density parameter becomes larger (see Fig.~\ref{fig:5cosmoPS}), and hence, according to the equation~\eqref{eq:defR*PL} the rescaled $R_{\ast}$ will decrease with the smoothing scale
and matter density. On the other hand, $R_{\ast}/\Rs$ does not vary with the dark energy EOS since it has no impact on the power spectrum slope. Consequently, when relying on the linear relation between $\Rex$ and $R_{\ast}$ (see Fig.~\ref{fig:Rstar_Rex}) the observed behaviors of $\Rex$ depicted in Figures~\ref{fig:Rex_cosmo} and~\ref{fig:Rex_Rs_fit} can be qualitatively explained with the prediction made for $R_{\ast}$ based on Gaussian random fields.

\subsection{Exclusion radius and dark energy EOS}\label{sec:results-w}

Let us now describe  how we can extract  information on the dark energy parameter using the exclusion radius,
even though the effect of dark energy on the measured exclusion radius is shown to be negligible.
Our strategy is based on the redshift invariance of the exclusion radius in comoving space. 

We follow the approach presented in \citet{ParkKim2010,blake+14,appleby+18}, where matter density and dark energy parameters were constrained to minimize the apparent redshift evolution of the genus of the underlying matter density field.

Let us recall the strategy. It relies on the fact that the redshift invariance of the measured genus amplitude is achieved only when the adopted cosmology for the redshift-distance relation matches the underlying true cosmology. Indeed, the genus amplitude, ${\cal A}$, obeys
\begin{equation}
{\cal A} \propto (\langle k^2\rangle)^{3/2}\propto \frac{1}{R_{0}^{3}}\,, \label{eq:defA}
\end{equation}
where
$R_{0}\equiv {\sigma_{0}}/{\sigma_{1}}$  characterizes the typical separation between zero crossing of a density field.
Hence $\cal A$ is expressed in terms of the ratio of moments defined by equation~\eqref{eq:defsig}.
Again, for scale-invariant power-spectra,  the genus amplitude $\cal A$ only depends on the effective power-index, $\ns$, as 
$\sqrt{2/(\ns+3)}$. Since the slope of a power spectrum changes with $R_{\rm s}$ (because the power spectrum is not scale invariant on the scales we are considering, the lower panel of Fig.~\ref{fig:5cosmoPS}), if an incorrect cosmology is assumed, the adopted smoothing scale effectively corresponds to a different length scale, hence  $\cal A$ is altered\footnote{In fact, genus-based estimators are  more sensitive 
to slope changes near $n_{\rm s}=-1$ because $R_0\propto1/\sqrt{\ns+3}$, whereas
$R_\ast\propto 1/\sqrt{\ns+5}$.}.

Similarly and importantly, if the wrong cosmology is adopted, we also expect to observe an apparent redshift evolution of $\Rex$ because it is linearly related to $R_\ast$ (as described in Appendix~\ref{sec:appendix-scaling}). We highlight that the measurements of exclusion radii in the evolved matter density field, which has deviated from its initial Gaussian nature, well align with the predictions made in the Gaussian random field for $R_{\ast}$. Therefore, in theory, it is indeed possible to estimate the deviation of the exclusion radius from the reference point as a function of redshift by calculating the effective smoothing scale for an adopted cosmological model.

\begin{figure}
\includegraphics[clip,width=1.1\columnwidth]{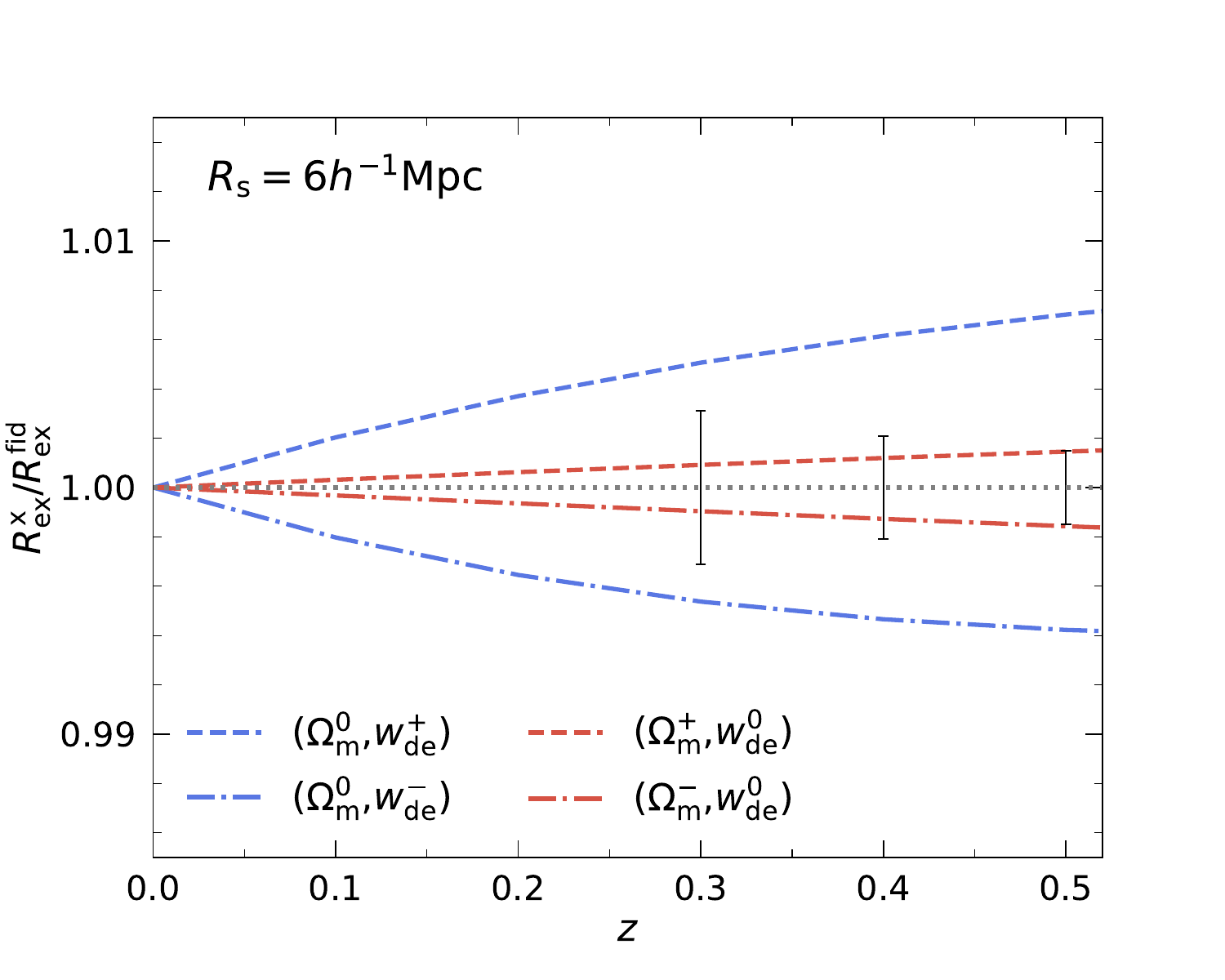}
    \caption{Expected redshift evolution of $\Rex$ relative to the true value when adopting trial cosmologies with different $\Om$ (red) or $w_{\rm de}$ (blue) for the redshift-distance relation. The exclusion radius for the true underlying cosmology is shown in a grey dotted line and vertical error bars at redshifts $z=0.3, 0.4$, and $0.5$ represent the estimated measurement uncertainties about the fiducial values, given the volumes surveyed up to those respective redshifts. Note that the true exclusion radius does not evolve with redshift.
    }    \label{fig:Rstar_z_exptected_R6}
\end{figure}

In Fig.~\ref{fig:Rstar_z_exptected_R6}, we show how the exclusion radius for a trial cosmology deviates from the fiducial case as a function of redshift. We observe a redshift evolution of the exclusion radius when the trial cosmology is inconsistent with the true underlying cosmology. For example, the exclusion radius monotonically grows with redshift when the matter density and dark energy parameters are larger than the true values, whereas it becomes smaller at higher redshifts in cosmologies with smaller matter density and dark energy parameters. We observe a larger departure from the reference for non-fiducial dark energy models. This reflects the fact that, at low $z$, the distance estimates are more strongly impacted by dark energy than matter density, in the redshift range of interest. Thus, the evolution of the apparent exclusion radius is more sensitive to the properties of dark energy.

We then calculate the potential measurement errors for an all-sky survey up to various redshifts to assess if it is possible to detect such redshift evolution of the exclusion radius given those survey volumes.
We base our error estimates on the standard error achieved from the simulation volume $V_{\rm sim}=1~h^{-3}{\rm Gpc}^{3}$. Let us assume that these errors reduce by $\sqrt{V_{\rm sim}/V_{\rm sur}}$, similar to the behavior of shot-noise, where $V_{\rm sur}$ is survey volume. For a survey with $V_{\rm sur}\approx2.5~h^{-3}{\rm Gpc}^{3}$, scanning up to $z=0.3$,  the departure of $w_{\rm de}$--shifted dark energy models from the fiducial cosmology can be detected approximately at $1.5\sigma$ significance level. When a survey extends to $z=0.5$ reaching $V_{\rm sur}\approx10~h^{-3}{\rm Gpc}^{3}$, the detection significances for the non-fiducial dark energy models can increase approximately to $3.8\sigma$, while those for $\Om$--shifted models eventually become marginal. 
We expect that one can more significantly detect the redshift variation of the exclusion radius when measuring it at higher redshifts. This is partly because the predicted amplitude of redshift variation gradually grows with redshift. In addition, even with a relatively narrow redshift span, the volume scanned through a non-full-sky survey targeting higher redshifts can be larger than the largest comoving volume considered here, for instance, $V_{\rm sur}=10~h^{-3}{\rm Gpc}^{3}$ up to $z\approx0.5$. Therefore, a higher significance detection should be available from surveys aiming at high redshifts. Note that we can also combine the exclusion radius measurements using four distinct types of cross-correlations, which should yield more robust measurements (although these measurements from the same density field are not strictly independent).
For completeness, we also show the expected redshift evolution of the exclusion radius for $R_{\rm s}=16\mpcph$ with the same mock surveys in Fig.~\ref{fig:Rstar_z_predicted_R16}. As expected, the error bars are larger.

\section{Conclusions \& Perspectives} \label{sec:conclusions}

This paper advocates that   the exclusion  radii, $\Rex$, define standard rulers on multiple scales:
on  linear scales, they can 
be computed  using Lagrangian theory \citep{BBKS,shim+21}, while on smaller non-linear scales they can be extracted from
cosmological simulations, as was done in this paper.
The value of $\Rex$ mainly constrains $\Om$ (Fig.~\ref{fig:rex_signi}) and may also be potentially used for measuring $\ns$ as can be done with the genus amplitude analysis \citep{appleby+20}.
Our results support that, at a given physical scale, the exclusion radius is redshift-invariant at first order, which can be leveraged to constrain the equation of state of dark energy $w_{\rm de}$ (Fig.~\ref{fig:Rstar_z_exptected_R6}). More precise results could be obtained by calibrating the scale-dependence and redshift evolution of the relation on $N$-body simulations.
We find that the redshift evolution and cosmology dependence of $\Rex$ can be remarkably captured using the locally-scale-invariant quantity $R_\ast$, which we can predict from linear theory.
This implies that the topology we measure at late time is already imprinted in the Gaussian initial state, and is not destroyed by gravitational collapse. Thus, the non-Gaussian final state mostly preserves its Gaussian initial topology.
We carried out a  naive error budget and found that 
the accuracy of the parameter estimation corresponds to a 4$\sigma$ detection of a change in matter content of 5\% in $1~h^{-3}{\rm Gpc}^{3}$, or roughly at 3.8$\sigma$ detection of 50\% change in the dark energy parameter using a full sky survey up to redshift 0.5. These numbers could be improved by performing a joint analysis of all exclusion
radii, accounting for the fact that they are not independent, or by exploring jointly different rarity thresholds.

At this stage, we assumed that the linear relation between $\Rex$ and $R_{\ast}$   given in Appendix~\ref{sec:appendix-scaling} holds for the range of redshifts considered; investigating its redshift dependence is of interest as it would allow us to mitigate
its residual effect (beyond the linear scaling shown in Fig.~\ref{fig:Rstar_Rex}) on smaller scales,
but is beyond the scope of the current analysis. Conversely, extending this  investigation to higher redshift would clearly 
be of interest given the depth of upcoming surveys\footnote{
\href{https://www.euclid-ec.org}{Euclid},
\href{http://www.lsst.org}{LSST},
\href{https://www.nasa.gov/content/goddard/nancy-grace-roman-space-telescope}{WFIRST},
\href{https://spherex.caltech.edu}{SphereX},
\href{https://ingconfluence.ing.iac.es:8444/confluence//display/WEAV/The+WEAVE+Project}{WEAVE},
\href{https://www.cfht.hawaii.edu/en/news/MSE-new/}{MSE},
\href{https://pfs.ipmu.jp}{PFS},
to name a few.}, and is also postponed to later work.

The accuracy of these standard rulers was assessed here using the multiverse set of simulations in a fairly idealized setting.
Our goal was to demonstrate the potential to build such rulers out of the two-point correlation functions of critical points, setting aside complications from completeness issues, tracer biases, SNR, and redshift space distortions.
The companion paper, \cite{Kraljic2022},  investigated how Lyman$-\alpha$ tomography can, in the context of the \weave mission \citep{WEAVE2012,WEAVE-QSO2016}, 
open access to the clustering of critical points and  the possibility of exclusion zone 
estimation. 
While the focus of this paper was on exclusion zones, \cite{Kraljic2022} showed that the position of the maxima of the cross-correlations of critical points of the same sign could also be estimated from tomography. Combining both approaches could potentially allow even tighter constraints to be set on cosmological parameters.
While more realistic, \cite{Kraljic2022}
still only involved mock data, and should of course be revisited with other tomographic 3D surveys 
such as PFS \citep{PFS2014}, MOSAIC on ELT \citep{MOSAIC2018}, MSE \citep{MSE2019},
but also with
spectroscopic or photometric redshifts galactic surveys, such as Euclid \citep{Euclid2011},
DESI \citep{DESI2016}, PFS, WFIRST \citep[Roman Space Telescope;][]{WFIRST2013}, or LSST \citep[Rubin Observatory;][]{LSST2019}. 

The  cosmology dependence of the clustering of critical point 
 should also be investigated in 2D maps such as weak lensing maps \citep{Euclid2011,KIDS2017}, line intensity maps \citep{CHIME2022} and photometric redshift surveys \citep{SphereX2014}.

It would finally be of interest to compute the auto-correlation function of saddle points in redshift space, which 
are less subject to finger of god effects given the density they probe.
This would allow for a more robust Alcock-Paczyński  test \citep{AP1979}, as one would expect that the overall motion of the local cosmic web does not impact such correlations.

\section*{Acknowledgements}
We thank C. Laigle for insightful conversations. We also thank the KITP for hosting the workshop \href{https://www.cosmicweb23.org}{`\emph{CosmicWeb23: connecting Galaxies to Cosmology at High and Low Redshift}'} during which this project was advanced. JS is supported by Academia Sinica Institute of Astronomy and Astrophysics, and thanks the Korea Institute for Advanced Studies for its hospitality when this paper was completed. SA is supported by an appointment to the JRG Program at the APCTP through the Science and Technology Promotion Fund and Lottery Fund of the Korean Government, and was also supported by the Korean Local Governments in Gyeongsangbuk-do Province and Pohang City. JK was supported by a KIAS Individual Grant (KG039603) via the Center for Advanced Computation at Korea Institute for Advanced Study. This work is partially supported by the National Science Foundation under Grant No. NSF PHY-1748958. We thank the Korea Institute for Advanced Study for providing computing resources (KIAS Center for Advanced Computation Linux Cluster System).

\section*{Data Availability}

The data underlying this article will be shared on reasonable request to the corresponding author.



\bibliographystyle{mnras}
\bibliography{reference} 

\begin{thebibliography}{}
\makeatletter
\relax
\def\mn@urlcharsother{\let\do\@makeother \do\$\do\&\do\#\do\^\do\_\do\%\do\~}
\def\mn@doi{\begingroup\mn@urlcharsother \@ifnextchar [ {\mn@doi@} {\mn@doi@[]}}
\def\mn@doi@[#1]#2{\def\@tempa{#1}\ifx\@tempa\@empty \href {http://dx.doi.org/#2} {doi:#2}\else \href {http://dx.doi.org/#2} {#1}\fi \endgroup}
\def\mn@eprint#1#2{\mn@eprint@#1:#2::\@nil}
\def\mn@eprint@arXiv#1{\href {http://arxiv.org/abs/#1} {{\tt arXiv:#1}}}
\def\mn@eprint@dblp#1{\href {http://dblp.uni-trier.de/rec/bibtex/#1.xml} {dblp:#1}}
\def\mn@eprint@#1:#2:#3:#4\@nil{\def\@tempa {#1}\def\@tempb {#2}\def\@tempc {#3}\ifx \@tempc \@empty \let \@tempc \@tempb \let \@tempb \@tempa \fi \ifx \@tempb \@empty \def\@tempb {arXiv}\fi \@ifundefined {mn@eprint@\@tempb}{\@tempb:\@tempc}{\expandafter \expandafter \csname mn@eprint@\@tempb\endcsname \expandafter{\@tempc}}}

\bibitem[\protect\citeauthoryear{{Alam} et~al.,}{{Alam} et~al.}{2021}]{2021PhRvD.103h3533A}
{Alam} S.,  et~al., 2021, \mn@doi [\prd] {10.1103/PhysRevD.103.083533}, \href {https://ui.adsabs.harvard.edu/abs/2021PhRvD.103h3533A} {103, 083533}

\bibitem[\protect\citeauthoryear{{Alcock} \& {Paczynski}}{{Alcock} \& {Paczynski}}{1979}]{AP1979}
{Alcock} C.,  {Paczynski} B.,  1979, \mn@doi [\nat] {10.1038/281358a0}, \href {https://ui.adsabs.harvard.edu/abs/1979Natur.281..358A} {281, 358}

\bibitem[\protect\citeauthoryear{{Appleby}, {Park}, {Hong}  \& {Kim}}{{Appleby} et~al.}{2018a}]{appleby+18}
{Appleby} S.,  {Park} C.,  {Hong} S.~E.,   {Kim} J.,  2018a, \mn@doi [\apj] {10.3847/1538-4357/aaa24f}, \href {https://ui.adsabs.harvard.edu/abs/2018ApJ...853...17A} {853, 17}

\bibitem[\protect\citeauthoryear{{Appleby}, {Chingangbam}, {Park}, {Hong}, {Kim}  \& {Ganesan}}{{Appleby} et~al.}{2018b}]{appleby+18minkowski_1}
{Appleby} S.,  {Chingangbam} P.,  {Park} C.,  {Hong} S.~E.,  {Kim} J.,   {Ganesan} V.,  2018b, \mn@doi [\apj] {10.3847/1538-4357/aabb53}, \href {https://ui.adsabs.harvard.edu/abs/2018ApJ...858...87A} {858, 87}

\bibitem[\protect\citeauthoryear{{Appleby}, {Park}, {Hong}, {Hwang}  \& {Kim}}{{Appleby} et~al.}{2020}]{appleby+20}
{Appleby} S.,  {Park} C.,  {Hong} S.~E.,  {Hwang} H.~S.,   {Kim} J.,  2020, \mn@doi [\apj] {10.3847/1538-4357/ab952e}, \href {https://ui.adsabs.harvard.edu/abs/2020ApJ...896..145A} {896, 145}

\bibitem[\protect\citeauthoryear{{Appleby}, {Park}, {Hong}, {Hwang}, {Kim}  \& {Tonegawa}}{{Appleby} et~al.}{2021}]{appleby+21}
{Appleby} S.,  {Park} C.,  {Hong} S.~E.,  {Hwang} H.~S.,  {Kim} J.,   {Tonegawa} M.,  2021, \mn@doi [\apj] {10.3847/1538-4357/abcebb}, \href {https://ui.adsabs.harvard.edu/abs/2021ApJ...907...75A} {907, 75}

\bibitem[\protect\citeauthoryear{{Appleby}, {Park}, {Pranav}, {Hong}, {Hwang}, {Kim}  \& {Buchert}}{{Appleby} et~al.}{2022}]{appleby+22}
{Appleby} S.,  {Park} C.,  {Pranav} P.,  {Hong} S.~E.,  {Hwang} H.~S.,  {Kim} J.,   {Buchert} T.,  2022, \mn@doi [\apj] {10.3847/1538-4357/ac562a}, \href {https://ui.adsabs.harvard.edu/abs/2022ApJ...928..108A} {928, 108}

\bibitem[\protect\citeauthoryear{{Armijo}, {Cai}, {Padilla}, {Li}  \& {Peacock}}{{Armijo} et~al.}{2018}]{armijo+18}
{Armijo} J.,  {Cai} Y.-C.,  {Padilla} N.,  {Li} B.,   {Peacock} J.~A.,  2018, \mn@doi [\mnras] {10.1093/mnras/sty1335}, \href {https://ui.adsabs.harvard.edu/abs/2018MNRAS.478.3627A} {478, 3627}

\bibitem[\protect\citeauthoryear{{Baldauf}, {Codis}, {Desjacques}  \& {Pichon}}{{Baldauf} et~al.}{2016}]{baldauf+16}
{Baldauf} T.,  {Codis} S.,  {Desjacques} V.,   {Pichon} C.,  2016, \mn@doi [\mnras] {10.1093/mnras/stv2973}, \href {https://ui.adsabs.harvard.edu/abs/2016MNRAS.456.3985B} {456, 3985}

\bibitem[\protect\citeauthoryear{{Bardeen}, {Bond}, {Kaiser}  \& {Szalay}}{{Bardeen} et~al.}{1986}]{BBKS}
{Bardeen} J.~M.,  {Bond} J.~R.,  {Kaiser} N.,   {Szalay} A.~S.,  1986, \apj, 304, 15

\bibitem[\protect\citeauthoryear{{Barthelemy}, {Codis}  \& {Bernardeau}}{{Barthelemy} et~al.}{2021}]{2021MNRAS.503.5204B}
{Barthelemy} A.,  {Codis} S.,   {Bernardeau} F.,  2021, \mn@doi [\mnras] {10.1093/mnras/stab818}, \href {https://ui.adsabs.harvard.edu/abs/2021MNRAS.503.5204B} {503, 5204}

\bibitem[\protect\citeauthoryear{{Bernardeau}}{{Bernardeau}}{1994}]{Bernardeau1994}
{Bernardeau} F.,  1994, \mn@doi [\apj] {10.1086/174620}, \href {https://ui.adsabs.harvard.edu/abs/1994ApJ...433....1B} {433, 1}

\bibitem[\protect\citeauthoryear{{Bernardeau} \& {Valageas}}{{Bernardeau} \& {Valageas}}{2000}]{2000A&A...364....1B}
{Bernardeau} F.,  {Valageas} P.,  2000, \mn@doi [\aap] {10.48550/arXiv.astro-ph/0006270}, \href {https://ui.adsabs.harvard.edu/abs/2000A&A...364....1B} {364, 1}

\bibitem[\protect\citeauthoryear{{Beutler} et~al.,}{{Beutler} et~al.}{2011}]{2011MNRAS.416.3017B}
{Beutler} F.,  et~al., 2011, \mn@doi [\mnras] {10.1111/j.1365-2966.2011.19250.x}, \href {https://ui.adsabs.harvard.edu/abs/2011MNRAS.416.3017B} {416, 3017}

\bibitem[\protect\citeauthoryear{{Beutler} et~al.,}{{Beutler} et~al.}{2017}]{beutler+17}
{Beutler} F.,  et~al., 2017, \mn@doi [\mnras] {10.1093/mnras/stw3298}, \href {https://ui.adsabs.harvard.edu/abs/2017MNRAS.466.2242B} {466, 2242}

\bibitem[\protect\citeauthoryear{{Blake} et~al.,}{{Blake} et~al.}{2011}]{blake+11}
{Blake} C.,  et~al., 2011, \mn@doi [\mnras] {10.1111/j.1365-2966.2011.19606.x}, \href {https://ui.adsabs.harvard.edu/abs/2011MNRAS.418.1725B} {418, 1725}

\bibitem[\protect\citeauthoryear{{Blake}, {James}  \& {Poole}}{{Blake} et~al.}{2014}]{blake+14}
{Blake} C.,  {James} J.~B.,   {Poole} G.~B.,  2014, \mn@doi [\mnras] {10.1093/mnras/stt2062}, \href {https://ui.adsabs.harvard.edu/abs/2014MNRAS.437.2488B} {437, 2488}

\bibitem[\protect\citeauthoryear{{Bond}, {Kofman}  \& {Pogosyan}}{{Bond} et~al.}{1996}]{bond+96}
{Bond} J.~R.,  {Kofman} L.,   {Pogosyan} D.,  1996, \mn@doi [\nat] {10.1038/380603a0}, \href {https://ui.adsabs.harvard.edu/abs/1996Natur.380..603B} {380, 603}

\bibitem[\protect\citeauthoryear{{Bouchet}, {Strauss}, {Davis}, {Fisher}, {Yahil}  \& {Huchra}}{{Bouchet} et~al.}{1993}]{Bouchet+1993}
{Bouchet} F.~R.,  {Strauss} M.~A.,  {Davis} M.,  {Fisher} K.~B.,  {Yahil} A.,   {Huchra} J.~P.,  1993, \mn@doi [\apj] {10.1086/173289}, \href {https://ui.adsabs.harvard.edu/abs/1993ApJ...417...36B} {417, 36}

\bibitem[\protect\citeauthoryear{{Boyle}, {Barthelemy}, {Codis}, {Habib}, {Uhlemann}  \& {Friedrich}}{{Boyle} et~al.}{2023}]{2023OJAp....6E..22B}
{Boyle} A.,  {Barthelemy} A.,  {Codis} S.,  {Habib} S.,  {Uhlemann} C.,   {Friedrich} O.,  2023, \mn@doi [The Open Journal of Astrophysics] {10.21105/astro.2212.10351}, \href {https://ui.adsabs.harvard.edu/abs/2023OJAp....6E..22B} {6, 22}

\bibitem[\protect\citeauthoryear{{CHIME Collaboration} et~al.,}{{CHIME Collaboration} et~al.}{2022}]{CHIME2022}
{CHIME Collaboration} et~al., 2022, \mn@doi [\apjs] {10.3847/1538-4365/ac6fd9}, \href {https://ui.adsabs.harvard.edu/abs/2022ApJS..261...29C} {261, 29}

\bibitem[\protect\citeauthoryear{{Cadiou}, {Pichon}, {Codis}, {Musso}, {Pogosyan}, {Dubois}, {Cardoso}  \& {Prunet}}{{Cadiou} et~al.}{2020}]{cadiou2020}
{Cadiou} C.,  {Pichon} C.,  {Codis} S.,  {Musso} M.,  {Pogosyan} D.,  {Dubois} Y.,  {Cardoso} J.~F.,   {Prunet} S.,  2020, \mn@doi [\mnras] {10.1093/mnras/staa1853}, \href {https://ui.adsabs.harvard.edu/abs/2020MNRAS.496.4787C} {496, 4787}

\bibitem[\protect\citeauthoryear{{Cappi} et~al.,}{{Cappi} et~al.}{2015}]{Cappi+2015}
{Cappi} A.,  et~al., 2015, \mn@doi [\aap] {10.1051/0004-6361/201525727}, \href {https://ui.adsabs.harvard.edu/abs/2015A&A...579A..70C} {579, A70}

\bibitem[\protect\citeauthoryear{{Chingangbam}, {Park}, {Yogendran}  \& {van de Weygaert}}{{Chingangbam} et~al.}{2012}]{chingangbam+12}
{Chingangbam} P.,  {Park} C.,  {Yogendran} K.~P.,   {van de Weygaert} R.,  2012, \mn@doi [\apj] {10.1088/0004-637X/755/2/122}, \href {https://ui.adsabs.harvard.edu/abs/2012ApJ...755..122C} {755, 122}

\bibitem[\protect\citeauthoryear{{Colombi}, {Pogosyan}  \& {Souradeep}}{{Colombi} et~al.}{2000}]{Colombi2000}
{Colombi} S.,  {Pogosyan} D.,   {Souradeep} T.,  2000, \mn@doi [\prl] {10.1103/PhysRevLett.85.5515}, \href {https://ui.adsabs.harvard.edu/abs/2000PhRvL..85.5515C} {85, 5515}

\bibitem[\protect\citeauthoryear{{Croton} et~al.,}{{Croton} et~al.}{2004}]{Croton+2004}
{Croton} D.~J.,  et~al., 2004, \mn@doi [\mnras] {10.1111/j.1365-2966.2004.08017.x}, \href {https://ui.adsabs.harvard.edu/abs/2004MNRAS.352.1232C} {352, 1232}

\bibitem[\protect\citeauthoryear{{DESI Collaboration} et~al.,}{{DESI Collaboration} et~al.}{2016}]{DESI2016}
{DESI Collaboration} et~al., 2016, arXiv e-prints, \href {https://ui.adsabs.harvard.edu/abs/2016arXiv161100036D} {p. arXiv:1611.00036}

\bibitem[\protect\citeauthoryear{{Da {\^A}ngela}, {Outram}  \& {Shanks}}{{Da {\^A}ngela} et~al.}{2005}]{2005MNRAS.361..879D}
{Da {\^A}ngela} J.,  {Outram} P.~J.,   {Shanks} T.,  2005, \mn@doi [\mnras] {10.1111/j.1365-2966.2005.09212.x}, \href {https://ui.adsabs.harvard.edu/abs/2005MNRAS.361..879D} {361, 879}

\bibitem[\protect\citeauthoryear{{Dalton} et~al.,}{{Dalton} et~al.}{2012}]{WEAVE2012}
{Dalton} G.,  et~al., 2012, in {McLean} I.~S.,  {Ramsay} S.~K.,   {Takami} H.,  eds,  Society of Photo-Optical Instrumentation Engineers (SPIE) Conference Series Vol. 8446, Ground-based and Airborne Instrumentation for Astronomy IV. p. 84460P, \mn@doi{10.1117/12.925950}

\bibitem[\protect\citeauthoryear{{Davis} \& {Peebles}}{{Davis} \& {Peebles}}{1983}]{1983ApJ...267..465D}
{Davis} M.,  {Peebles} P.~J.~E.,  1983, \mn@doi [\apj] {10.1086/160884}, \href {https://ui.adsabs.harvard.edu/abs/1983ApJ...267..465D} {267, 465}

\bibitem[\protect\citeauthoryear{{Dawson} et~al.,}{{Dawson} et~al.}{2013}]{dawson+13}
{Dawson} K.~S.,  et~al., 2013, \mn@doi [\aj] {10.1088/0004-6256/145/1/10}, \href {https://ui.adsabs.harvard.edu/abs/2013AJ....145...10D} {145, 10}

\bibitem[\protect\citeauthoryear{{Dong}, {Park}, {Hong}, {Kim}, {Hwang}, {Park}  \& {appleby+18minkeby}}{{Dong} et~al.}{2023}]{dong+23}
{Dong} F.,  {Park} C.,  {Hong} S.~E.,  {Kim} J.,  {Hwang} H.~S.,  {Park} H.,   {appleby+18minkeby} S.,  2023, \mn@doi [\apj] {10.3847/1538-4357/acd185}, \href {https://ui.adsabs.harvard.edu/abs/2023ApJ...953...98D} {953, 98}

\bibitem[\protect\citeauthoryear{{Dor{\'e}} et~al.,}{{Dor{\'e}} et~al.}{2014}]{SphereX2014}
{Dor{\'e}} O.,  et~al., 2014, \mn@doi [arXiv e-prints] {10.48550/arXiv.1412.4872}, \href {https://ui.adsabs.harvard.edu/abs/2014arXiv1412.4872D} {p. arXiv:1412.4872}

\bibitem[\protect\citeauthoryear{{Dubinski}, {Kim}, {Park}  \& {Humble}}{{Dubinski} et~al.}{2004}]{dubinski+04}
{Dubinski} J.,  {Kim} J.,  {Park} C.,   {Humble} R.,  2004, \mn@doi [\na] {10.1016/j.newast.2003.08.002}, \href {https://ui.adsabs.harvard.edu/abs/2004NewA....9..111D} {9, 111}

\bibitem[\protect\citeauthoryear{{Dunkley} et~al.,}{{Dunkley} et~al.}{2009}]{dunkley+09}
{Dunkley} J.,  et~al., 2009, \mn@doi [\apjs] {10.1088/0067-0049/180/2/306}, \href {https://ui.adsabs.harvard.edu/abs/2009ApJS..180..306D} {180, 306}

\bibitem[\protect\citeauthoryear{Eisenstein et~al.,}{Eisenstein et~al.}{2005}]{eisenstein_DetectionBaryonAcoustic_2005a}
Eisenstein D.~J.,  et~al., 2005, \mn@doi [The Astrophysical Journal] {10.1086/466512}, 633, 560

\bibitem[\protect\citeauthoryear{{Feldbrugge}, {van Engelen}, {van de Weygaert}, {Pranav}  \& {Vegter}}{{Feldbrugge} et~al.}{2019}]{Feldbrugge+2019}
{Feldbrugge} J.,  {van Engelen} M.,  {van de Weygaert} R.,  {Pranav} P.,   {Vegter} G.,  2019, \mn@doi [\jcap] {10.1088/1475-7516/2019/09/052}, \href {https://ui.adsabs.harvard.edu/abs/2019JCAP...09..052F} {2019, 052}

\bibitem[\protect\citeauthoryear{{Fry}}{{Fry}}{1985}]{Fry1985}
{Fry} J.~N.,  1985, \mn@doi [\apj] {10.1086/162859}, \href {https://ui.adsabs.harvard.edu/abs/1985ApJ...289...10F} {289, 10}

\bibitem[\protect\citeauthoryear{{Gay}, {Pichon}  \& {Pogosyan}}{{Gay} et~al.}{2012}]{Gay2012}
{Gay} C.,  {Pichon} C.,   {Pogosyan} D.,  2012, \prd, 85, 023011

\bibitem[\protect\citeauthoryear{{Gott}, {Melott}  \& {Dickinson}}{{Gott} et~al.}{1986}]{GottMellotDickinson1986}
{Gott} J.~Richard I.,  {Melott} A.~L.,   {Dickinson} M.,  1986, \mn@doi [\apj] {10.1086/164347}, \href {https://ui.adsabs.harvard.edu/abs/1986ApJ...306..341G} {306, 341}

\bibitem[\protect\citeauthoryear{{Goyal}, {Chingangbam}  \& {appleby+18minkeby}}{{Goyal} et~al.}{2020}]{goyal+20minkowski}
{Goyal} P.,  {Chingangbam} P.,   {appleby+18minkeby} S.,  2020, \mn@doi [\jcap] {10.1088/1475-7516/2020/02/020}, \href {https://ui.adsabs.harvard.edu/abs/2020JCAP...02..020G} {2020, 020}

\bibitem[\protect\citeauthoryear{{Hikage} et~al.,}{{Hikage} et~al.}{2003}]{hikage+03}
{Hikage} C.,  et~al., 2003, \mn@doi [\pasj] {10.1093/pasj/55.5.911}, \href {https://ui.adsabs.harvard.edu/abs/2003PASJ...55..911H} {55, 911}

\bibitem[\protect\citeauthoryear{{Hildebrandt} et~al.,}{{Hildebrandt} et~al.}{2017}]{KIDS2017}
{Hildebrandt} H.,  et~al., 2017, \mn@doi [\mnras] {10.1093/mnras/stw2805}, \href {https://ui.adsabs.harvard.edu/abs/2017MNRAS.465.1454H} {465, 1454}

\bibitem[\protect\citeauthoryear{{Hinshaw}, {Banday}, {Bennett}, {Gorski}  \& {Kogut}}{{Hinshaw} et~al.}{1995}]{Hinshaw+1995}
{Hinshaw} G.,  {Banday} A.~J.,  {Bennett} C.~L.,  {Gorski} K.~M.,   {Kogut} A.,  1995, \mn@doi [\apjl] {10.1086/187932}, \href {https://ui.adsabs.harvard.edu/abs/1995ApJ...446L..67H} {446, L67}

\bibitem[\protect\citeauthoryear{{Ivezi{\'c}} et~al.,}{{Ivezi{\'c}} et~al.}{2019}]{LSST2019}
{Ivezi{\'c}} {\v{Z}}.,  et~al., 2019, \mn@doi [\apj] {10.3847/1538-4357/ab042c}, \href {https://ui.adsabs.harvard.edu/abs/2019ApJ...873..111I} {873, 111}

\bibitem[\protect\citeauthoryear{{James}, {Colless}, {Lewis}  \& {Peacock}}{{James} et~al.}{2009}]{BerianJames+2009}
{James} J.~B.,  {Colless} M.,  {Lewis} G.~F.,   {Peacock} J.~A.,  2009, \mn@doi [\mnras] {10.1111/j.1365-2966.2008.14358.x}, \href {https://ui.adsabs.harvard.edu/abs/2009MNRAS.394..454J} {394, 454}

\bibitem[\protect\citeauthoryear{{Jenkins}}{{Jenkins}}{2010}]{jenkins+10}
{Jenkins} A.,  2010, \mn@doi [\mnras] {10.1111/j.1365-2966.2010.16259.x}, \href {https://ui.adsabs.harvard.edu/abs/2010MNRAS.403.1859J} {403, 1859}

\bibitem[\protect\citeauthoryear{{Junaid} \& {Pogosyan}}{{Junaid} \& {Pogosyan}}{2015}]{junaid+15}
{Junaid} M.,  {Pogosyan} D.,  2015, \mn@doi [\prd] {10.1103/PhysRevD.92.043505}, \href {https://ui.adsabs.harvard.edu/abs/2015PhRvD..92d3505J} {92, 043505}

\bibitem[\protect\citeauthoryear{{Kaiser}}{{Kaiser}}{1984}]{Kaiser1984}
{Kaiser} N.,  1984, \apjl, 284, L9

\bibitem[\protect\citeauthoryear{{Kraljic} et~al.,}{{Kraljic} et~al.}{2022}]{Kraljic2022}
{Kraljic} K.,  et~al., 2022, \mn@doi [\mnras] {10.1093/mnras/stac1409}, \href {https://ui.adsabs.harvard.edu/abs/2022MNRAS.514.1359K} {514, 1359}

\bibitem[\protect\citeauthoryear{{Laureijs} et~al.,}{{Laureijs} et~al.}{2011}]{Euclid2011}
{Laureijs} R.,  et~al., 2011, \mn@doi [arXiv e-prints] {10.48550/arXiv.1110.3193}, \href {https://ui.adsabs.harvard.edu/abs/2011arXiv1110.3193L} {p. arXiv:1110.3193}

\bibitem[\protect\citeauthoryear{{Li}, {Park}, {Sabiu}, {Park}, {Weinberg}, {Schneider}, {Kim}  \& {Hong}}{{Li} et~al.}{2016}]{li+16}
{Li} X.-D.,  {Park} C.,  {Sabiu} C.~G.,  {Park} H.,  {Weinberg} D.~H.,  {Schneider} D.~P.,  {Kim} J.,   {Hong} S.~E.,  2016, \mn@doi [\apj] {10.3847/0004-637X/832/2/103}, \href {https://ui.adsabs.harvard.edu/abs/2016ApJ...832..103L} {832, 103}

\bibitem[\protect\citeauthoryear{{Li} et~al.,}{{Li} et~al.}{2018}]{li+18}
{Li} X.-D.,  et~al., 2018, \mn@doi [\apj] {10.3847/1538-4357/aab42e}, \href {https://ui.adsabs.harvard.edu/abs/2018ApJ...856...88L} {856, 88}

\bibitem[\protect\citeauthoryear{{Lumsden}, {Heavens}  \& {Peacock}}{{Lumsden} et~al.}{1989}]{lumsden+89}
{Lumsden} S.~L.,  {Heavens} A.~F.,   {Peacock} J.~A.,  1989, \mn@doi [\mnras] {10.1093/mnras/238.2.293}, \href {https://ui.adsabs.harvard.edu/abs/1989MNRAS.238..293L} {238, 293}

\bibitem[\protect\citeauthoryear{{Mar{\'\i}n} et~al.,}{{Mar{\'\i}n} et~al.}{2013}]{marin+13}
{Mar{\'\i}n} F.~A.,  et~al., 2013, \mn@doi [\mnras] {10.1093/mnras/stt520}, \href {https://ui.adsabs.harvard.edu/abs/2013MNRAS.432.2654M} {432, 2654}

\bibitem[\protect\citeauthoryear{{Marinoni} \& {Buzzi}}{{Marinoni} \& {Buzzi}}{2010}]{2010Natur.468..539M}
{Marinoni} C.,  {Buzzi} A.,  2010, \mn@doi [\nat] {10.1038/nature09577}, \href {https://ui.adsabs.harvard.edu/abs/2010Natur.468..539M} {468, 539}

\bibitem[\protect\citeauthoryear{{Marques} et~al.,}{{Marques} et~al.}{2023}]{marques+23}
{Marques} G.~A.,  et~al., 2023, \mn@doi [arXiv e-prints] {10.48550/arXiv.2308.10866}, \href {https://ui.adsabs.harvard.edu/abs/2023arXiv230810866M} {p. arXiv:2308.10866}

\bibitem[\protect\citeauthoryear{{Massara}, {Villaescusa-Navarro}, {Ho}, {Dalal}  \& {Spergel}}{{Massara} et~al.}{2021}]{massara+21}
{Massara} E.,  {Villaescusa-Navarro} F.,  {Ho} S.,  {Dalal} N.,   {Spergel} D.~N.,  2021, \mn@doi [\prl] {10.1103/PhysRevLett.126.011301}, \href {https://ui.adsabs.harvard.edu/abs/2021PhRvL.126a1301M} {126, 011301}

\bibitem[\protect\citeauthoryear{{Massara} et~al.,}{{Massara} et~al.}{2023}]{massara+23}
{Massara} E.,  et~al., 2023, \mn@doi [\apj] {10.3847/1538-4357/acd44d}, \href {https://ui.adsabs.harvard.edu/abs/2023ApJ...951...70M} {951, 70}

\bibitem[\protect\citeauthoryear{{Mecke} \& {Wagner}}{{Mecke} \& {Wagner}}{1991}]{MeckeWagner1991}
{Mecke} K.~R.,  {Wagner} H.,  1991, \mn@doi [Journal of Statistical Physics] {10.1007/BF01048319}, \href {https://ui.adsabs.harvard.edu/abs/1991JSP....64..843M} {64, 843}

\bibitem[\protect\citeauthoryear{{Mecke}, {Buchert}  \& {Wagner}}{{Mecke} et~al.}{1994}]{MeckeBuchertWagner1994}
{Mecke} K.~R.,  {Buchert} T.,   {Wagner} H.,  1994, \aap, \href {https://ui.adsabs.harvard.edu/abs/1994A&A...288..697M} {288, 697}

\bibitem[\protect\citeauthoryear{{Melott}, {Cohen}, {Hamilton}, {Gott}  \& {Weinberg}}{{Melott} et~al.}{1989}]{Melott+1989}
{Melott} A.~L.,  {Cohen} A.~P.,  {Hamilton} A. J.~S.,  {Gott} J.~Richard I.,   {Weinberg} D.~H.,  1989, \mn@doi [\apj] {10.1086/167935}, \href {https://ui.adsabs.harvard.edu/abs/1989ApJ...345..618M} {345, 618}

\bibitem[\protect\citeauthoryear{Milnor}{Milnor}{1963}]{milnor1963morse}
Milnor J.,  1963, Morse theory.
Princeton University Press, Princeton, N.J

\bibitem[\protect\citeauthoryear{{Mo} \& {White}}{{Mo} \& {White}}{1996}]{mo&white96}
{Mo} H.~J.,  {White} S.~D.~M.,  1996, \mn@doi [\mnras] {10.1093/mnras/282.2.347}, \href {https://ui.adsabs.harvard.edu/abs/1996MNRAS.282..347M} {282, 347}

\bibitem[\protect\citeauthoryear{{Natoli} et~al.,}{{Natoli} et~al.}{2010}]{natoli+10}
{Natoli} P.,  et~al., 2010, \mn@doi [\mnras] {10.1111/j.1365-2966.2010.17228.x}, \href {https://ui.adsabs.harvard.edu/abs/2010MNRAS.408.1658N} {408, 1658}

\bibitem[\protect\citeauthoryear{{Neveux} et~al.,}{{Neveux} et~al.}{2020}]{neveux+20}
{Neveux} R.,  et~al., 2020, \mn@doi [\mnras] {10.1093/mnras/staa2780}, \href {https://ui.adsabs.harvard.edu/abs/2020MNRAS.499..210N} {499, 210}

\bibitem[\protect\citeauthoryear{{Nichol} et~al.,}{{Nichol} et~al.}{2006}]{Nichol+2006}
{Nichol} R.~C.,  et~al., 2006, \mn@doi [\mnras] {10.1111/j.1365-2966.2006.10239.x}, \href {https://ui.adsabs.harvard.edu/abs/2006MNRAS.368.1507N} {368, 1507}

\bibitem[\protect\citeauthoryear{{Okumura}, {Matsubara}, {Eisenstein}, {Kayo}, {Hikage}, {Szalay}  \& {Schneider}}{{Okumura} et~al.}{2008}]{okumura+08}
{Okumura} T.,  {Matsubara} T.,  {Eisenstein} D.~J.,  {Kayo} I.,  {Hikage} C.,  {Szalay} A.~S.,   {Schneider} D.~P.,  2008, \mn@doi [\apj] {10.1086/528951}, \href {https://ui.adsabs.harvard.edu/abs/2008ApJ...676..889O} {676, 889}

\bibitem[\protect\citeauthoryear{{Okumura} et~al.,}{{Okumura} et~al.}{2016}]{okumra+16}
{Okumura} T.,  et~al., 2016, \mn@doi [\pasj] {10.1093/pasj/psw029}, \href {https://ui.adsabs.harvard.edu/abs/2016PASJ...68...38O} {68, 38}

\bibitem[\protect\citeauthoryear{{Park} \& {Gott}}{{Park} \& {Gott}}{1991}]{ParkGott1991}
{Park} C.,  {Gott} J.~R. I.,  1991, \mn@doi [\apj] {10.1086/170445}, \href {https://ui.adsabs.harvard.edu/abs/1991ApJ...378..457P} {378, 457}

\bibitem[\protect\citeauthoryear{{Park} \& {Kim}}{{Park} \& {Kim}}{2010}]{ParkKim2010}
{Park} C.,  {Kim} Y.-R.,  2010, \mn@doi [\apjl] {10.1088/2041-8205/715/2/L185}, \href {https://ui.adsabs.harvard.edu/abs/2010ApJ...715L.185P} {715, L185}

\bibitem[\protect\citeauthoryear{{Park}, {Gott}  \& {Choi}}{{Park} et~al.}{2001}]{Park+2001}
{Park} C.,  {Gott} J.~Richard I.,   {Choi} Y.~J.,  2001, \mn@doi [\apj] {10.1086/320640}, \href {https://ui.adsabs.harvard.edu/abs/2001ApJ...553...33P} {553, 33}

\bibitem[\protect\citeauthoryear{{Park} et~al.,}{{Park} et~al.}{2013}]{Park+2013}
{Park} C.,  et~al., 2013, \mn@doi [Journal of Korean Astronomical Society] {10.5303/JKAS.2013.46.3.125}, \href {https://ui.adsabs.harvard.edu/abs/2013JKAS...46..125P} {46, 125}

\bibitem[\protect\citeauthoryear{{Park}, {Park}, {Sabiu}, {Li}, {Hong}, {Kim}, {Tonegawa}  \& {Zheng}}{{Park} et~al.}{2019}]{park+19}
{Park} H.,  {Park} C.,  {Sabiu} C.~G.,  {Li} X.-d.,  {Hong} S.~E.,  {Kim} J.,  {Tonegawa} M.,   {Zheng} Y.,  2019, \mn@doi [\apj] {10.3847/1538-4357/ab2da1}, \href {https://ui.adsabs.harvard.edu/abs/2019ApJ...881..146P} {881, 146}

\bibitem[\protect\citeauthoryear{{Peebles} \& {Groth}}{{Peebles} \& {Groth}}{1975}]{Peebles+1975}
{Peebles} P.~J.~E.,  {Groth} E.~J.,  1975, \mn@doi [\apj] {10.1086/153390}, \href {https://ui.adsabs.harvard.edu/abs/1975ApJ...196....1P} {196, 1}

\bibitem[\protect\citeauthoryear{{Percival}, {Cole}, {Eisenstein}, {Nichol}, {Peacock}, {Pope}  \& {Szalay}}{{Percival} et~al.}{2007}]{2007MNRAS.381.1053P}
{Percival} W.~J.,  {Cole} S.,  {Eisenstein} D.~J.,  {Nichol} R.~C.,  {Peacock} J.~A.,  {Pope} A.~C.,   {Szalay} A.~S.,  2007, \mn@doi [\mnras] {10.1111/j.1365-2966.2007.12268.x}, \href {https://ui.adsabs.harvard.edu/abs/2007MNRAS.381.1053P} {381, 1053}

\bibitem[\protect\citeauthoryear{{Philcox}}{{Philcox}}{2022}]{philcox22}
{Philcox} O. H.~E.,  2022, \mn@doi [\prd] {10.1103/PhysRevD.106.063501}, \href {https://ui.adsabs.harvard.edu/abs/2022PhRvD.106f3501P} {106, 063501}

\bibitem[\protect\citeauthoryear{{Pieri} et~al.,}{{Pieri} et~al.}{2016}]{WEAVE-QSO2016}
{Pieri} M.~M.,  et~al., 2016, in {Reyl{\'e}} C.,  {Richard} J.,  {Cambr{\'e}sy} L.,  {Deleuil} M.,  {P{\'e}contal} E.,  {Tresse} L.,   {Vauglin} I.,  eds, SF2A-2016: Proceedings of the Annual meeting of the French Society of Astronomy and Astrophysics. pp 259--266 (\mn@eprint {arXiv} {1611.09388})

\bibitem[\protect\citeauthoryear{{Pogosyan}, {Pichon}, {Gay}, {Prunet}, {Cardoso}, {Sousbie}  \& {Colombi}}{{Pogosyan} et~al.}{2009}]{pogosyan+09}
{Pogosyan} D.,  {Pichon} C.,  {Gay} C.,  {Prunet} S.,  {Cardoso} J.~F.,  {Sousbie} T.,   {Colombi} S.,  2009, \mn@doi [\mnras] {10.1111/j.1365-2966.2009.14753.x}, \href {https://ui.adsabs.harvard.edu/abs/2009MNRAS.396..635P} {396, 635}

\bibitem[\protect\citeauthoryear{{Pranav}, {Edelsbrunner}, {van de Weygaert}, {Vegter}, {Kerber}, {Jones}  \& {Wintraecken}}{{Pranav} et~al.}{2017}]{Pranav+2017}
{Pranav} P.,  {Edelsbrunner} H.,  {van de Weygaert} R.,  {Vegter} G.,  {Kerber} M.,  {Jones} B. J.~T.,   {Wintraecken} M.,  2017, \mn@doi [\mnras] {10.1093/mnras/stw2862}, \href {https://ui.adsabs.harvard.edu/abs/2017MNRAS.465.4281P} {465, 4281}

\bibitem[\protect\citeauthoryear{{Pranav} et~al.,}{{Pranav} et~al.}{2019}]{Pranav+2019}
{Pranav} P.,  et~al., 2019, \mn@doi [\mnras] {10.1093/mnras/stz541}, \href {https://ui.adsabs.harvard.edu/abs/2019MNRAS.485.4167P} {485, 4167}

\bibitem[\protect\citeauthoryear{{Puech} et~al.,}{{Puech} et~al.}{2018}]{MOSAIC2018}
{Puech} M.,  et~al., 2018, in {Evans} C.~J.,  {Simard} L.,   {Takami} H.,  eds,  Society of Photo-Optical Instrumentation Engineers (SPIE) Conference Series Vol. 10702, Ground-based and Airborne Instrumentation for Astronomy VII. p. 107028R (\mn@eprint {arXiv} {1806.03296}), \mn@doi{10.1117/12.2311292}

\bibitem[\protect\citeauthoryear{{Sabiu}, {Hoyle}, {Kim}  \& {Li}}{{Sabiu} et~al.}{2019}]{sabiu+19}
{Sabiu} C.~G.,  {Hoyle} B.,  {Kim} J.,   {Li} X.-D.,  2019, \mn@doi [\apjs] {10.3847/1538-4365/ab22b5}, \href {https://ui.adsabs.harvard.edu/abs/2019ApJS..242...29S} {242, 29}

\bibitem[\protect\citeauthoryear{{Satpathy}, {A C Croft}, {Ho}  \& {Li}}{{Satpathy} et~al.}{2019}]{satpathy+19}
{Satpathy} S.,  {A C Croft} R.,  {Ho} S.,   {Li} B.,  2019, \mn@doi [\mnras] {10.1093/mnras/stz009}, \href {https://ui.adsabs.harvard.edu/abs/2019MNRAS.484.2148S} {484, 2148}

\bibitem[\protect\citeauthoryear{{Schmalzing} \& {Buchert}}{{Schmalzing} \& {Buchert}}{1997}]{SchmalzingBuchert1997}
{Schmalzing} J.,  {Buchert} T.,  1997, \mn@doi [\apjl] {10.1086/310680}, \href {https://ui.adsabs.harvard.edu/abs/1997ApJ...482L...1S} {482, L1}

\bibitem[\protect\citeauthoryear{{Sefusatti} \& {Vernizzi}}{{Sefusatti} \& {Vernizzi}}{2011}]{sefusatti&vernizzi11}
{Sefusatti} E.,  {Vernizzi} F.,  2011, \mn@doi [\jcap] {10.1088/1475-7516/2011/03/047}, \href {https://ui.adsabs.harvard.edu/abs/2011JCAP...03..047S} {2011, 047}

\bibitem[\protect\citeauthoryear{{Shandarin}}{{Shandarin}}{1983}]{Shandarin1983}
{Shandarin} S.~F.,  1983, Soviet Astronomy Letters, \href {https://ui.adsabs.harvard.edu/abs/1983SvAL....9..104S} {9, 104}

\bibitem[\protect\citeauthoryear{{Shandarin} \& {Yess}}{{Shandarin} \& {Yess}}{1998}]{ShandarinYess1998}
{Shandarin} S.~F.,  {Yess} C.,  1998, \mn@doi [\apj] {10.1086/306135}, \href {https://ui.adsabs.harvard.edu/abs/1998ApJ...505...12S} {505, 12}

\bibitem[\protect\citeauthoryear{{Sheth} \& {Lemson}}{{Sheth} \& {Lemson}}{1999}]{sheth&lemson99}
{Sheth} R.~K.,  {Lemson} G.,  1999, \mn@doi [\mnras] {10.1046/j.1365-8711.1999.02378.x}, \href {https://ui.adsabs.harvard.edu/abs/1999MNRAS.304..767S} {304, 767}

\bibitem[\protect\citeauthoryear{{Shim}, {Codis}, {Pichon}, {Pogosyan}  \& {Cadiou}}{{Shim} et~al.}{2021}]{shim+21}
{Shim} J.,  {Codis} S.,  {Pichon} C.,  {Pogosyan} D.,   {Cadiou} C.,  2021, \mn@doi [\mnras] {10.1093/mnras/stab263}, \href {https://ui.adsabs.harvard.edu/abs/2021MNRAS.502.3885S} {502, 3885}

\bibitem[\protect\citeauthoryear{{Slepian} et~al.,}{{Slepian} et~al.}{2017}]{slepian+17}
{Slepian} Z.,  et~al., 2017, \mn@doi [\mnras] {10.1093/mnras/stx488}, \href {https://ui.adsabs.harvard.edu/abs/2017MNRAS.469.1738S} {469, 1738}

\bibitem[\protect\citeauthoryear{{Sousbie}, {Pichon}, {Colombi}, {Novikov}  \& {Pogosyan}}{{Sousbie} et~al.}{2008}]{sousbie+08}
{Sousbie} T.,  {Pichon} C.,  {Colombi} S.,  {Novikov} D.,   {Pogosyan} D.,  2008, \mn@doi [\mnras] {10.1111/j.1365-2966.2007.12685.x}, \href {https://ui.adsabs.harvard.edu/abs/2008MNRAS.383.1655S} {383, 1655}

\bibitem[\protect\citeauthoryear{{Sousbie}, {Colombi}  \& {Pichon}}{{Sousbie} et~al.}{2009}]{sousbie+09}
{Sousbie} T.,  {Colombi} S.,   {Pichon} C.,  2009, \mn@doi [\mnras] {10.1111/j.1365-2966.2008.14244.x}, \href {https://ui.adsabs.harvard.edu/abs/2009MNRAS.393..457S} {393, 457}

\bibitem[\protect\citeauthoryear{{Sousbie}, {Pichon}  \& {Kawahara}}{{Sousbie} et~al.}{2011}]{Sousbie2011-2}
{Sousbie} T.,  {Pichon} C.,   {Kawahara} H.,  2011, \mn@doi [\mnras] {10.1111/j.1365-2966.2011.18395.x}, \href {https://ui.adsabs.harvard.edu/abs/2011MNRAS.414..384S} {414, 384}

\bibitem[\protect\citeauthoryear{{Spergel} et~al.,}{{Spergel} et~al.}{2013}]{WFIRST2013}
{Spergel} D.,  et~al., 2013, \mn@doi [arXiv e-prints] {10.48550/arXiv.1305.5422}, \href {https://ui.adsabs.harvard.edu/abs/2013arXiv1305.5422S} {p. arXiv:1305.5422}

\bibitem[\protect\citeauthoryear{{Sugiyama} et~al.,}{{Sugiyama} et~al.}{2023}]{sugiyama+23}
{Sugiyama} N.~S.,  et~al., 2023, \mn@doi [\mnras] {10.1093/mnras/stad1935}, \href {https://ui.adsabs.harvard.edu/abs/2023MNRAS.524.1651S} {524, 1651}

\bibitem[\protect\citeauthoryear{{Takada} et~al.,}{{Takada} et~al.}{2014}]{PFS2014}
{Takada} M.,  et~al., 2014, \mn@doi [\pasj] {10.1093/pasj/pst019}, \href {https://ui.adsabs.harvard.edu/abs/2014PASJ...66R...1T} {66, R1}

\bibitem[\protect\citeauthoryear{{The MSE Science Team} et~al.,}{{The MSE Science Team} et~al.}{2019}]{MSE2019}
{The MSE Science Team} et~al., 2019, \mn@doi [arXiv e-prints] {10.48550/arXiv.1904.04907}, \href {https://ui.adsabs.harvard.edu/abs/2019arXiv190404907T} {p. arXiv:1904.04907}

\bibitem[\protect\citeauthoryear{{Tonegawa}, {Park}, {Zheng}, {Park}, {Hong}, {Hwang}  \& {Kim}}{{Tonegawa} et~al.}{2020}]{tonegawa+20}
{Tonegawa} M.,  {Park} C.,  {Zheng} Y.,  {Park} H.,  {Hong} S.~E.,  {Hwang} H.~S.,   {Kim} J.,  2020, \mn@doi [\apj] {10.3847/1538-4357/ab95ff}, \href {https://ui.adsabs.harvard.edu/abs/2020ApJ...897...17T} {897, 17}

\bibitem[\protect\citeauthoryear{{Uhlemann}, {Codis}, {Pichon}, {Bernardeau}  \& {Reimberg}}{{Uhlemann} et~al.}{2016}]{uhlemann+16}
{Uhlemann} C.,  {Codis} S.,  {Pichon} C.,  {Bernardeau} F.,   {Reimberg} P.,  2016, \mn@doi [\mnras] {10.1093/mnras/stw1074}, \href {https://ui.adsabs.harvard.edu/abs/2016MNRAS.460.1529U} {460, 1529}

\bibitem[\protect\citeauthoryear{{Uhlemann}, {Codis}, {Kim}, {Pichon}, {Bernardeau}, {Pogosyan}, {Park}  \& {L'Huillier}}{{Uhlemann} et~al.}{2017}]{Uhlemann2017}
{Uhlemann} C.,  {Codis} S.,  {Kim} J.,  {Pichon} C.,  {Bernardeau} F.,  {Pogosyan} D.,  {Park} C.,   {L'Huillier} B.,  2017, \mn@doi [\mnras] {10.1093/mnras/stw3221}, \href {https://ui.adsabs.harvard.edu/abs/2017MNRAS.466.2067U} {466, 2067}

\bibitem[\protect\citeauthoryear{{Uhlemann}, {Friedrich}, {Villaescusa-Navarro}, {Banerjee}  \& {Codis}}{{Uhlemann} et~al.}{2020}]{2020MNRAS.495.4006U}
{Uhlemann} C.,  {Friedrich} O.,  {Villaescusa-Navarro} F.,  {Banerjee} A.,   {Codis} S.,  2020, \mn@doi [\mnras] {10.1093/mnras/staa1155}, \href {https://ui.adsabs.harvard.edu/abs/2020MNRAS.495.4006U} {495, 4006}

\bibitem[\protect\citeauthoryear{{Van de Weygaert} et~al.,}{{Van de Weygaert} et~al.}{2011}]{vandeWeygaert+2011}
{Van de Weygaert} R.,  et~al., 2011, \mn@doi [arXiv e-prints] {10.48550/arXiv.1110.5528}, \href {https://ui.adsabs.harvard.edu/abs/2011arXiv1110.5528V} {p. arXiv:1110.5528}

\bibitem[\protect\citeauthoryear{{White} \& {Padmanabhan}}{{White} \& {Padmanabhan}}{2009}]{2009MNRAS.395.2381W}
{White} M.,  {Padmanabhan} N.,  2009, \mn@doi [\mnras] {10.1111/j.1365-2966.2009.14732.x}, \href {https://ui.adsabs.harvard.edu/abs/2009MNRAS.395.2381W} {395, 2381}

\bibitem[\protect\citeauthoryear{{Xu}, {Jing}, {Zhao}  \& {Cuesta}}{{Xu} et~al.}{2023}]{xu+23}
{Xu} K.,  {Jing} Y.~P.,  {Zhao} G.-B.,   {Cuesta} A.~J.,  2023, \mn@doi [Nature Astronomy] {10.1038/s41550-023-02035-4}, \href {https://ui.adsabs.harvard.edu/abs/2023NatAs...7.1259X} {7, 1259}

\bibitem[\protect\citeauthoryear{{Yess}, {Shandarin}  \& {Fisher}}{{Yess} et~al.}{1997}]{Yess+1997}
{Yess} C.,  {Shandarin} S.~F.,   {Fisher} K.~B.,  1997, \mn@doi [\apj] {10.1086/303496}, \href {https://ui.adsabs.harvard.edu/abs/1997ApJ...474..553Y} {474, 553}

\bibitem[\protect\citeauthoryear{{Zhang}, {Cheng}  \& {Chu}}{{Zhang} et~al.}{2018}]{zhang+18}
{Zhang} J.,  {Cheng} D.,   {Chu} M.-C.,  2018, \mn@doi [\prd] {10.1103/PhysRevD.97.023534}, \href {https://ui.adsabs.harvard.edu/abs/2018PhRvD..97b3534Z} {97, 023534}

\bibitem[\protect\citeauthoryear{{Zhang}, {Huang}  \& {Li}}{{Zhang} et~al.}{2019}]{zhang+19}
{Zhang} X.,  {Huang} Q.-G.,   {Li} X.-D.,  2019, \mn@doi [\mnras] {10.1093/mnras/sty3191}, \href {https://ui.adsabs.harvard.edu/abs/2019MNRAS.483.1655Z} {483, 1655}

\makeatother
\end{thebibliography}



\appendix

\section{probing dark energy EOS on larger scales}\label{sec:appendix-16}

\begin{figure}
\includegraphics[clip,width=\columnwidth]{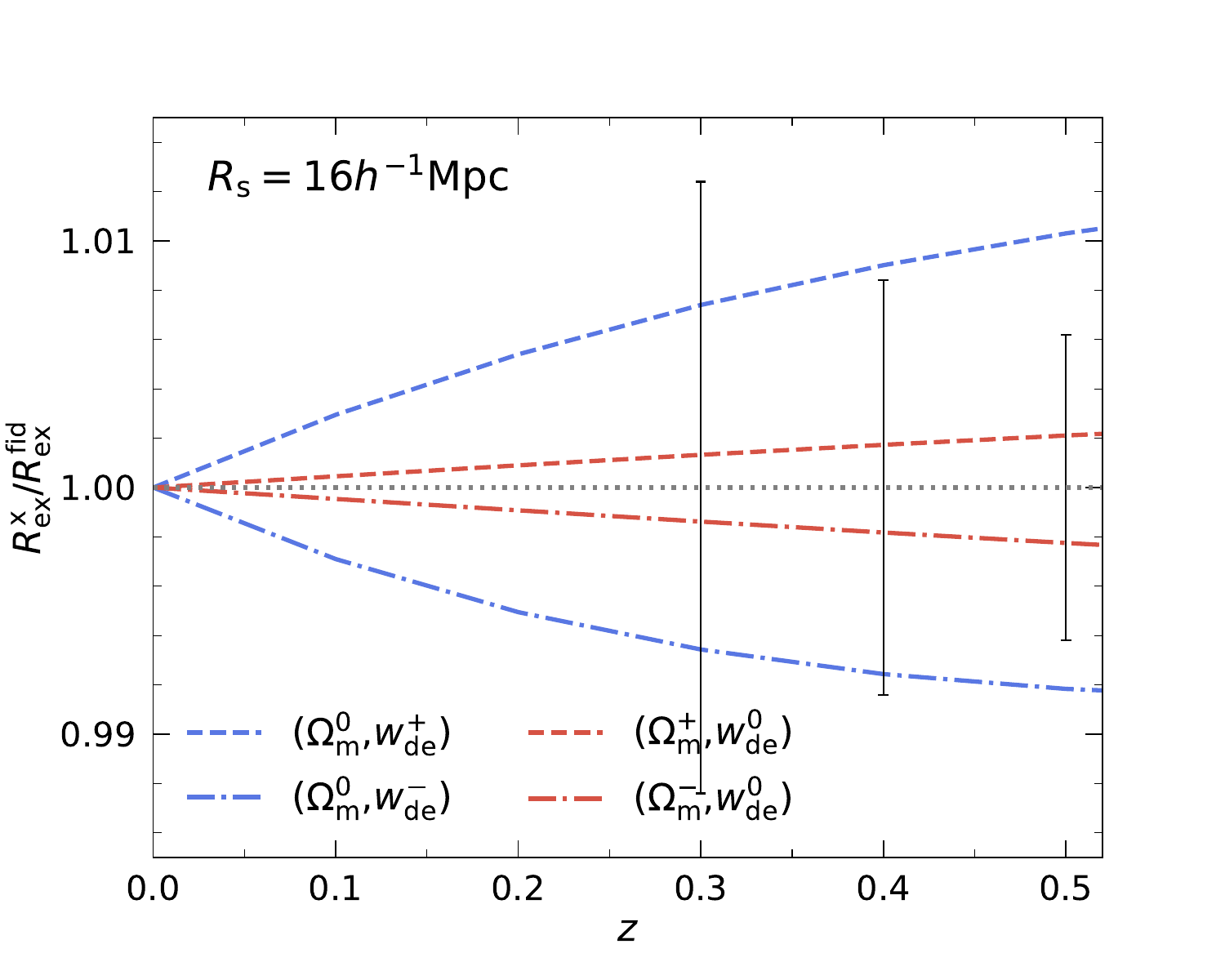}
\caption{Same as Fig.~\ref{fig:Rstar_z_exptected_R6} but for a smoothing scale $R_{\rm s}=16\mpcph$. Note that the deviation and measurement errors increase with smoothing scale. For example, the deviation and error are larger by factors of 1.4 and 3.5 for ($\Om^{0}, w_{\rm de}^{+}$) cosmology (blue dashed) at $z=0.5$.}
\label{fig:Rstar_z_predicted_R16}
\end{figure}
Fig.~\ref{fig:Rstar_z_predicted_R16} reproduces Fig.~\ref{fig:Rstar_z_exptected_R6} for a different smoothing scale,
 $R_{\rm s}=16\mpcph$. As expected, as the number of critical points decreases the error 
 bars for measuring $w_{\rm de}$ increases comparatively.

\section{Scaling relations }\label{sec:appendix-scaling}
Let us first describe how the shape of the matter power spectrum depends on cosmological parameters. Subsequently, we will elucidate the relation between $\Rex$ and $R_{\ast}$ in various cosmological models.

\begin{figure}
\includegraphics[clip,width=\columnwidth]{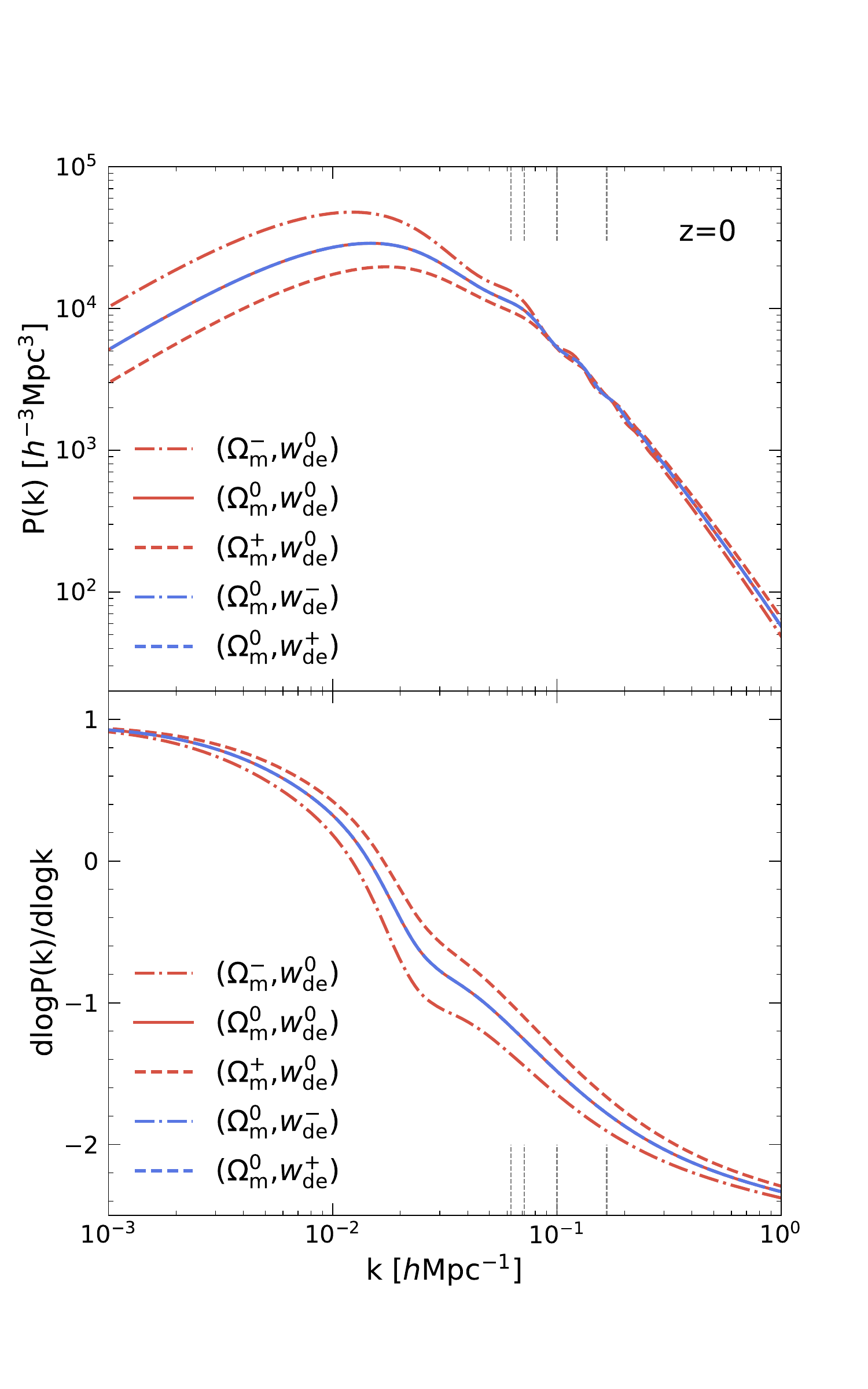}
    \caption{Linear matter power spectra (upper panel) and their slope (lower panel) for five cosmologies considered. From left to right, the vertical ticks located on $k=1/\Rs$ (grey dashed) mark wavenumbers corresponding to the Gaussian smoothing scales $R_{\rm s}=16, 14, 10$ and $6 \mpcph$, respectively. All power spectra have the identical normalization amplitude $\sigma_8$. Note that the power spectrum slope difference becomes larger on smaller wavenumbers than the smoothing lengths we consider. This implies that one can better constrain $\Om$ with a larger smoothing scale.}
    \label{fig:5cosmoPS}
\end{figure}

The linear matter power spectra for the cosmologies considered are depicted in Fig.~\ref{fig:5cosmoPS}. Primarily, the shape of the matter power spectrum varies with $\Om$, while its dependence on the dark energy equation of state $w_{\rm de}$ is minimal. The $\Om$-dependence of the power spectrum is attributed to the shift in the scale of matter-radiation equality as the matter density parameter changes.

\begin{figure}
\includegraphics[clip,width=\columnwidth]{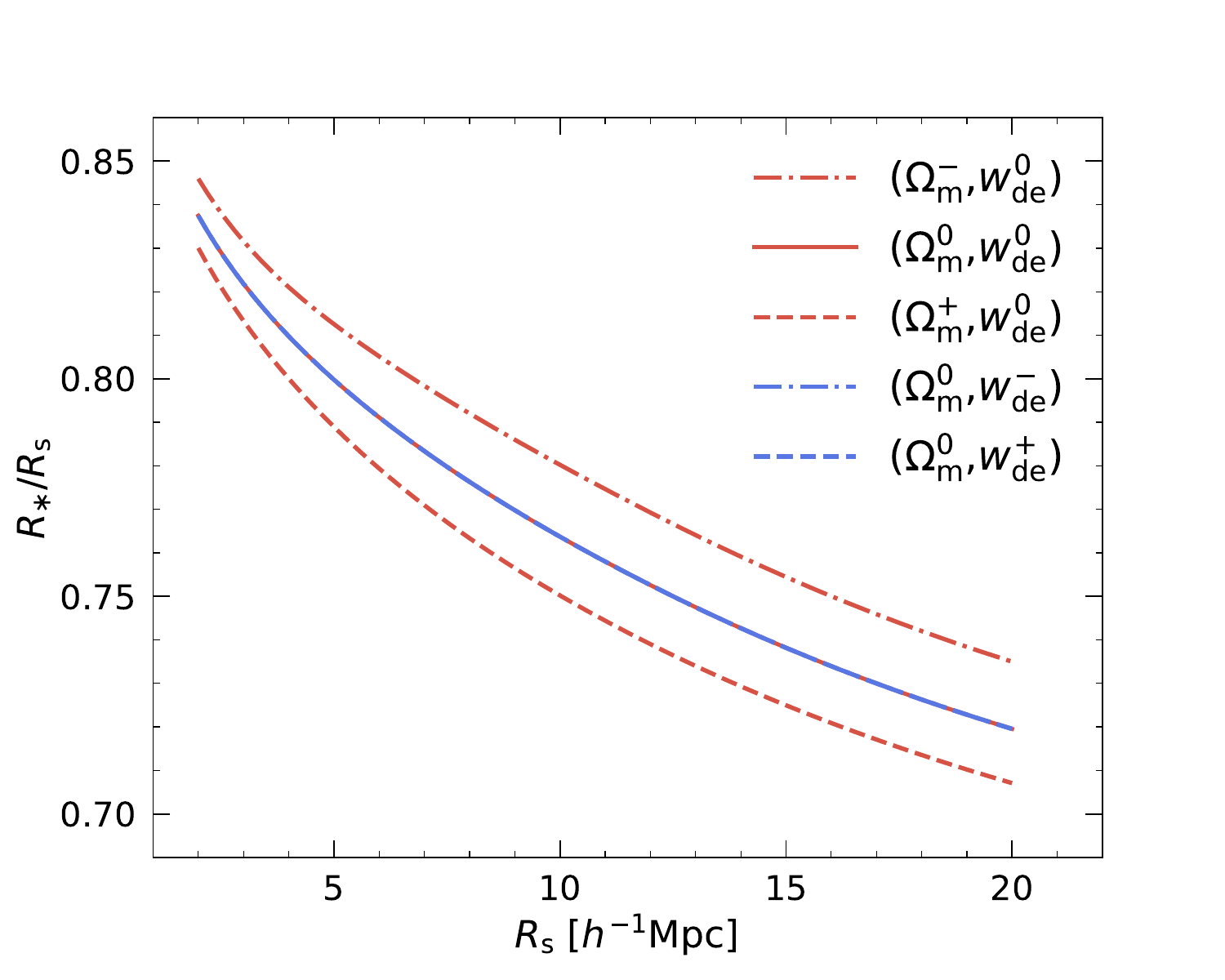}
\caption{Gaussian prediction for normalized $R_{\ast}$ as a function of smoothing scale $R_{\rm s}$ for five different cosmologies. $\Om$--shifted and $w_{\rm de}$--shifted cosmologies are shown in red and blue lines, respectively. Note that the predictions for the $w_{\rm de}$--shifted models (blue) perfectly overlap with the fiducial case (red solid). Note also that the moment ratios of a given power spectrum are redshift independent.}
\label{fig:Rstar_Rs_theory}
\end{figure}

In Fig.~\ref{fig:Rstar_Rs_theory}, we present the moment ratio $R_{\ast}$ as a function of the smoothing scale $R_{\rm s}$, based on the matter power spectra shown in Fig.~\ref{fig:5cosmoPS}. The normalized $R_{\ast}$ decreases either with increasing smoothing scale or larger matter density.
However, changing the dark energy EOS exhibits a minimal impact on $R_{\ast}$. As $R_{\ast}$ probes the power spectrum slope, $n_s$ at that scale (Equation~\eqref{eq:defR*PL}), we can interpret its matter density dependence as follows. Because the power spectrum amplitude $\sigma_8$ is fixed across different cosmologies, a power spectrum with larger $\Om$ has less large- and more small-scale power, as depicted in Fig.~\ref{fig:5cosmoPS}. Hence, increasing $\Om$ results in a shallower power spectrum slope on scales corresponding to the smoothing lengths considered in this analysis. Thus, $R_{\ast}$, tracing the slope, decreases with increasing $\Om$.
It is worth noting that the amplitude of the power spectrum grows over time. The rate of growth is influenced by both $\Om$ and $w_{\rm de}$. However, $R_{\ast}$ is expected to remain nearly constant with redshifts given that the growth factors in the moment ratio essentially cancel out.

\begin{figure*}
\includegraphics[clip,width=1.6\columnwidth]{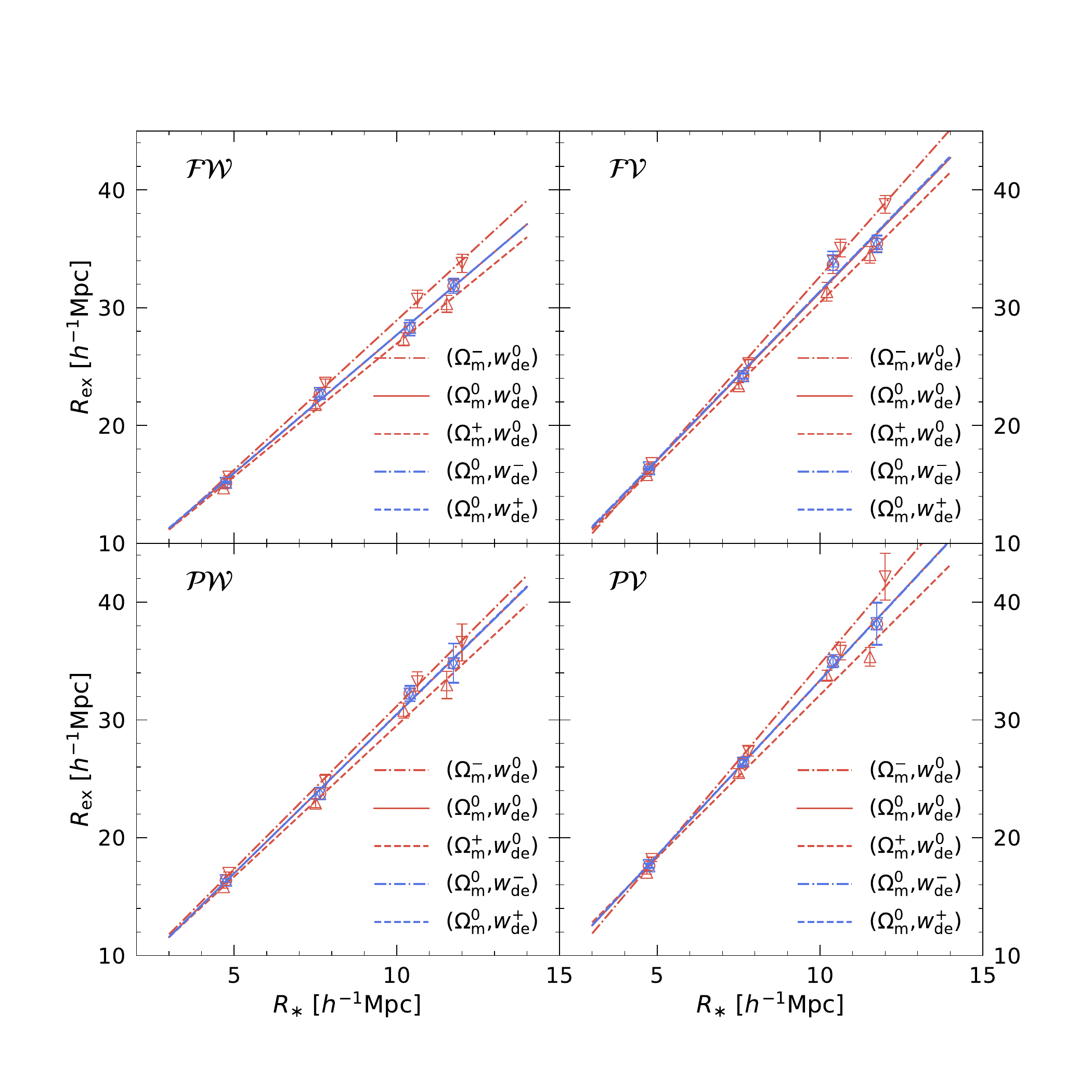}\caption{Relation between the measured exclusion radius $\Rex$ and the predicted $R_{*}$ for four different cross-correlation functions. Symbols from left to right correspond to the measurements and Gaussian predictions at smoothing scales $R_{\rm s}=6, 10, 14$, and $16 \mpcph$. Best linear fits for the symbols are shown in lines and the coefficients of the fits are given in Table~\ref{tab:coeffi}. Again, error bars represent the standard errors of the mean.}
\label{fig:Rstar_Rex}
\end{figure*}

With both measurements and predictions available, Fig.~\ref{fig:Rstar_Rex} illustrates the relation between the measured $\Rex$ for four distinct cross-correlations and the predicted $R_{\ast}$ in the Gaussian random field limit. The symbols denote measurements at smoothing scales of $R_{\rm s}=6, 10, 14$ and 16$\mpcph$, while the lines represent their respective best fits. Standard errors of the mean are shown with the error bars. The relation between $\Rex$ and $R_{\ast}$ can be effectively characterized by a linear relation, 
\begin{equation}
    R_{\rm ex}=a_{1}R_{\ast}+a_{2},
\end{equation}
with its coefficients detailed in Table~\ref{tab:coeffi}. The slope of the linear scaling relation shows its dependence on the matter density parameter but remains unaffected by changes in the dark energy EOS. Notably, in all cross-correlations, the slope becomes steeper with a smaller $\Om$.

\begin{table}
\centering
\caption{Slope, $a_{1}$, and intercept, $a_{2}$, of the linear scaling relations between $\Rex$ and $R_{\ast}$ for $\mathcal{FW}$, $\mathcal{FV}$, $\mathcal{PW}$, and $\mathcal{PV}$ for the cosmologies considered.}
\label{tab:coeffi}	
    \begin{tabular}{cccccc} 
    \hline\hline
    cosmology & & $\mathcal{FW}$ & $\mathcal{FV}$ & $\mathcal{PW}$ & $\mathcal{PV}$\\
    \hline
($\Om^{-}$, $w_{\rm de}^{0}$) & $a_1$ & 2.54 & 3.12 & 2.77 & 3.27\\
 & $a_2$ & 3.53 & 1.49 & 3.49 & 2.06\\
 \hline

($\Om^{0}, w_{\rm de}^{0}$) & $a_1$ & 2.35 & 2.85 & 2.70 & 2.97\\
& $a_2$ & 4.20 & 2.79 & 3.46 & 3.63\\
\hline

($\Om^{+}, w_{\rm de}^{0}$) & $a_1$ & 2.26 & 2.76 & 2.57 & 2.76\\
& $a_2$ & 4.38 & 2.87 & 3.85 & 4.54\\
\hline

($\Om^{0}, w_{\rm de}^{-}$) & $a_1$ & 2.35 & 2.85 & 2.70 & 2.97\\
& $a_2$ & 4.21 & 2.88 & 3.48 & 3.64\\
\hline

($\Om^{0}, w_{\rm de}^{+}$) & $a_1$ & 2.34 & 2.87 & 2.71 & 2.97\\
& $a_2$ & 4.28 & 2.73 & 3.42 & 3.67\\		
    \hline
    \end{tabular}
\end{table}

\bsp	
\label{lastpage}
\end{document}